\documentclass[12pt]{article}
\newlength{\abstractwidth}
\renewcommand{\thefootnote}{\fnsymbol{footnote}}
\renewcommand{\thanks}[1]{\footnote{#1}} 
\newcommand{\starttext}{
\setcounter{footnote}{0}
\renewcommand{\thefootnote}{\arabic{footnote}}}

\newcommand{\be}{\begin{equation}}
\newcommand{\bea}{\begin{eqnarray}}
\newcommand{\eea}{\end{eqnarray}}
\newcommand{\ee}{\end{equation}}
\newcommand{\no}{\nonumber}

\renewcommand{\>}{\rangle}
\def\ba{\begin{eqnarray}}
\def\ea{\end{eqnarray}}

\def\Tr{{\rm Tr}}
\def\det{{\rm det}}
\def\Det{{\rm Det}}
\def\half{ {1\over 2}}
\def\quar{{1 \over 4}}
\def\12{{\scriptstyle {1 \over 2}}}
\def\p{\partial}
\def\tet{\vartheta}
\def\ep{{\varepsilon}}
\def\C{{\cal C}}
\def\F{{\cal F}}
\def\H{{\cal H}}

\def\s2{\sqrt{2}}
\def\bZ{{\bf Z}}
\def\cS{{\cal S}}
\def\cZ{{\cal Z}}
\def\bR{{\bf R}}
\def\lbar{\ell \! \bar {} \, }
\begin{document}
\starttext 
\baselineskip=18pt 
\setcounter{footnote}{0}

\begin{flushright}
UCLA/04/TEP/01 \\
Columbia/Math/04 \\
\end{flushright}

\bigskip

\begin{center}

{ \Large \bf ON THE CONSTRUCTION OF ASYMMETRIC}
\medskip
{ \Large \bf ORBIFOLD MODELS}
\footnote{Research supported in part by a Grant-in-Aid from the Ministry
of Education, Science, Sports and Culture and grants from Keio University 
(K.A.),  and by  National Science Foundation
grants PHY-01-40151 (E.D.) and DMS-02-45371 (D.P.).}

\bigskip\bigskip

{\large \bf Kenichiro Aoki$^a$, Eric D'Hoker$^b$ and D.H. Phong$^c$}

\bigskip

$^a$ {\sl Hiyoshi Department of Physics} \\
{\sl Keio University, Yokohama 223-8521,  Japan} \\
$^b$ {\sl Department of Physics and Astronomy }\\
{\sl University of California, Los Angeles, CA 90095, USA} \\
$^c$ {\sl Department of Mathematics} \\
{\sl Columbia University, New York, NY 10027, USA}

\end{center}

\bigskip\bigskip

\begin{abstract}

\bigskip

Various asymmetric orbifold models based on chiral shifts and chiral
reflections are investigated. Special attention is devoted to 
1the consistency of the models with two fundamental principles  
for asymmetric orbifolds : modular invariance and the existence of
a proper Hilbert space formulation for states and operators.
The interplay between these two principles is non-trivial.
It is shown, for example, that their simultaneous requirement
forces the order of a chiral reflection to be 4, instead of the naive 2.
A careful explicit construction is given of the associated one-loop partition 
functions. At higher loops, the partition functions of asymmetric orbifolds
are built from the chiral blocks of associated symmetric orbifolds, 
whose pairings are determined by degenerations to one-loop.
\end{abstract}

\newpage

\baselineskip=15pt
\setcounter{equation}{0}
\setcounter{footnote}{0}

\section{Introduction}
\setcounter{equation}{0}

String theories are built of independent left and right movers on the 
worldsheet, and both Type II \cite{typeII} and Heterotic \cite{heterotic} 
superstring theories 
are inherently chiral on the worldsheet. As such, they admit asymmetric 
compactifications for left and right degrees  of freedom. This fact is basic to 
the 
construction of the Heterotic string \cite{heterotic} and to other models such 
as the 
Narain compactification \cite{narain}. 
Compactifications by orbifolds of flat space-time
(\cite{orbifolds, dfms}; for reviews see  \cite{g88,dms,orbrev}), are 
particularly
important because their conformal field theories are explicitly solvable
from free field theories via group theoretic methods, and yet their 
space-time properties are non-trivial. 

\medskip

For symmetric orbifolds,
the orbifold group acts identically on left and right movers. 
The associated worldsheet conformal field theory admits a consistent 
construction
equivalently via functional integral  or operator methods \cite{vafa,DVV,dvvv}, 
and 
the functional integral formulation naturally guarantees modular invariance.

\medskip

For asymmetric orbifolds, the orbifold group acts differently on left 
and right movers. This circumstance renders functional integral formulations
problematic. Therefore,   
modular invariance and the validity of a Hilbert space interpretation cannot be 
taken for granted. 
Some general principles for compactifications on asymmetric orbifolds have 
been formulated in \cite{nsv}, but relatively few
examples have been studied explicitly (see for example \cite{extras}. 
(At special compactification radii, asymmetric 
orbifold conformal field theories may be reformulated in terms of free fermionic 
degrees of freedom alone. The action of the orbifold group on chiral fermions 
is well understood and systematic studies of asymmetric free fermion models 
are available in \cite{freefermions1,freefermions2}.) 

\medskip

Asymmetric orbifolds are clearly needed when orbifolding the Heterotic string 
(for example, see \cite{polchinski}).
More recently, asymmetric orbifold models  for Type II strings
generated by chiral reflections
and shifts were also proposed \cite{KKS1,KKS2, KKS4,KKS3}. 
These ``Kachru-Kumar-Silverstein" (KKS)
models are of particular interest, 
since they exhibit space-time supersymmetry breaking 
and vanishing cosmological constant at one-loop order.
Initially, there were hopes that the cosmological constant
would continue to vanish to two loops 
\cite{KKS1,KKS2} (see also \cite{ksiz}), via an independent
cancellation of the sums over spin structures for left and right movers,
but this is now known not to be the case \cite{adp}. 

\medskip

In the present paper, we take the opportunity of the study of the
KKS models to examine the principles of the
construction of asymmetric orbifold models based on chiral
shifts and twists in greater detail,
and to add to the hitherto relatively short list of examples
worked out explicitly. 
In view of the methods of chiral splitting \cite{dp87,dp88,dp89} (see also 
\cite{vv2,vv1}) and two-loop 
superstring perturbation theory developed over the past few years in 
\cite{I,II,III,IV,V} (see also \cite{zhu,zhu1}), 
it is natural to postulate that the partition function for
an asymmetric orbifold is obtained by suitably pairing the chiral blocks of 
symmetric orbifold theories.  Thus, the main problem
of asymmetry is the determination of the corresponding pairing matrix
(see \cite{adp} and \S 6). 
It can be determined in principle by degenerations to one-loop. 
Thus, {\sl the key problem} reduces to
finding all the one-loop traces 
$\cZ^g{}_h={\rm Tr}_{{\cal H}_h}(g q^{L_0}\bar q^{\bar L_0})$,
where $g,h$ runs over all the elements of the orbifold group,
and ${\cal H}_h$ is the Hilbert space of the sector twisted by $h$.

\medskip

Even for the relatively simple asymmetric orbifold
models based on chiral shifts and twists, 
there are important subtleties in the construction:
(1) chiral operators $g,h$ are defined only up to phases, 
which can affect their orders and change the whole structure
of the theory; (2) it is unclear how to incorporate in
the construction of the Hilbert space ${\cal H}_h$
for the asymmetric theory the ground state degeneracies of the
corresponding symmetric orbifold theories; (3) one has to
extend to ${\cal H}_h$ the operator $g$ which was originally 
defined only on the untwisted Hilbert space. 

\medskip
We find that for the asymmetric orbifold models based
on chiral shifts and twists, the combined requirements
of modular covariance and Hilbert space interpretation constrain
the orders, and hence the phases, of the elements
of the orbifold group. The degeneracies in the Hilbert spaces
for the twisted asymmetric theory can be handled by suitable
selection rules on the larger set of ground states coming from
the symmetric theory. And upon this construction, the operators
of the theory admit consistent extensions to the twisted sectors.
It can be hoped that similar considerations will apply to
more general asymmetric orbifold models.

\subsection{Principles of Orbifold Constructions}

Let $G$ be an orbifold group acting on a flat torus  ${\bf T}^n$.
The orbifold  group $G$ will be taken to be finite and Abelian for 
simplicity.\footnote{For a more detailed summary of 
symmetric and asymmetric orbifold constructions, 
see \cite{adp}. The orbifold construction may be carried out
either by coseting flat Euclidean space ${\bf R}^n$ by a full orbifold
group including the action of translations, or by coseting the flat torus
${\bf T}^n$ by the point group $\bar P_G$. For symmetric orbifolds,
these procedures are equivalent, as is demonstrated in some simple
cases in Appendix B. For asymmetric orbifolds, the coseting procedure 
starting from the torus is taken as a definition.} 
The following principles will be taken as the starting point
for orbifold constructions on the worldsheet of a torus with modulus $\tau$;

\begin{itemize} 

\item {\bf The existence of a consistent Hilbert space formulation}

The partition function $Z_G$ of the $G$-orbifold theory is given by 
a summation over {\sl partition traces} $\cZ ^g {}_h$,
\bea
\label{ZG}
Z_G (\tau) ={1\over |G|}\sum_{g,h\in G} \cZ^g{}_h(\tau),
\hskip .8in
 \cZ^g{}_h(\tau) \equiv \Tr_{\H_h}(g\,q^{L_0}\bar q^{\bar L_0}),
\eea
where $ q \equiv \exp \{2\pi i\tau \}$ and $|G|$ is the order of $G$.  
The key assumption in this principle is that a consistent Hilbert space
formulation exists in which $\H_h$ is the Hilbert space of  the sector 
twisted by $h$, and the group elements $g \in G$ have a consistent 
operator realization in each of these twisted Hilbert spaces. 

\item {\bf Modular Covariance of partition traces}

Modular invariance of the partition function $Z_G$ is guaranteed
by the stronger condition of modular covariance of each of the partition traces,
\bea
\label{modularcovariance}
\cZ^g{}_h(\tau+1)&=&\cZ^{gh^{-1}}{}_h(\tau)
\no \\
\cZ^g{}_h(-1/\tau)&=&\cZ^{h^{-1}}{}_g(\tau)
\eea
or more generally,
\bea
\label{genmodcov}
\cZ^g{}_h \left ( {a \tau + b \over c \tau +d} \right )
=
\cZ^{g^d h^{-b}}{}_{g^{-c} h^a} (\tau)
\hskip .8in 
\left ( \matrix{a & b \cr c & d \cr} \right ) \in SL(2,\bZ)
\eea
Modular covariance of the partition traces implies that 
$\cZ^g{}_h(\tau+n_h)=\cZ^g{}_h(\tau)$ for any element $h$ 
of order $n_h$. This relation, in turn, is equivalent to
the familiar requirement of level matching,
\be
\label{levelmatching}
L_0-\bar L_0 \in {1\over n_h}\,{\bf Z}
\ee
when acting on the Hilbert space $\H_h$ of states twisted by an 
element $h$ of order $n_h$. Finally, the transformation in
(\ref{genmodcov}) given by $a=d=-1$ and $b=c=0$ 
belongs to the center of $SL(2,\bZ)$,
reverses the orientation of both $A$ and $B$ cycles, 
and corresponds to charge conjugation symmetry.

\end{itemize}

When the orbifold group $G$ acts on genuine well-defined fields, 
and a functional integral formulation is available, the above 
principles may be deduced from the functional integral formulation
using standard quantum field theory methods. 
In particular, a simple change of variables in the functional 
integral will imply the the modular covariance relations for the partition 
traces
expressed above.

\medskip

For general asymmetric orbifolds, a proper functional integral formulation 
may be lacking and  it may not be possible to derive the above relations 
from first principles. In such theories, the above principles will simply
be postulated as necessary and sufficient conditions for the 
existence of physically viable asymmetric orbifold models.  
It has been argued in \cite{vafa} 
that level matching suffices to insure the full modular covariance 
(\ref{modularcovariance}) of the theory. This condition is not always 
sufficient to guarantee the full modular covariance of the partition
traces. In this paper some examples will be presented where
this lack of modular covariance conflicts with the existence of a 
proper Hilbert space interpretation.

\subsection{Algorithm for the construction of asymmetric orbifolds} 

In this paper, asymmetric orbifold models based on chiral shifts and reflections
(such as arise in the constructions of \cite{KKS1}) are examined in detail. 
Special attention is devoted to a proper construction of states and operators 
in the untwisted and twisted sectors of Hilbert space and to the modular
covariance of the partition traces. Models with only chiral shifts are 
constructed
first and shown to satisfy modular covariance, at least in some critical 
dimensions.
In models involving chiral reflections, difficulties are found with modular 
covariance, even though the level matching condition (\ref{levelmatching}) 
is satisfied . These problems are traced
to the precise definition of the chiral reflection operators in Hilbert space.
In particular, it is shown that a chiral reflection of order 2 leads to 
conflicts
with the modular transformation $\tau \to -1/\tau$, but that a chiral reflection
of order 4 is consistent with modular invariance. 

\medskip

Generally,  in models satisfying level matching, it had been expected that 
the modular orbit method would yield all partition traces $\cZ^g{}_h(\tau)$ 
from applying the modular covariance condition (\ref{modularcovariance}) 
to the partition traces in the untwisted sector $\cZ ^g {}_1(\tau)$. 
The models with chiral reflections examined here show that the method must be 
applied with some care: modular transformations may not generate all partition 
traces from the untwisted sector, and even when they do, the objects they 
generate may not have a proper interpretation as partition traces in a twisted 
sector. 

\medskip

While our discussions later in the paper will be in terms of specific
examples based on shifts and twists, the methods developed there
may be recast in terms of a recursive algorithm, 
which we shall now summarize.

\begin{enumerate}

\item 
It is assumed that all the group elements $g\in G$ in the orbifold group have 
well-defined operator realizations on the untwisted Hilbert space $\H_1$.
The partition traces $\cZ ^g {}_1$ in the untwisted 
sector for all $g\in G$ are then well-defined and may be calculated.

\item
Modular transformations applied to the partition traces $\cZ ^h {}_1$ 
yield $\cZ ^1 {}_h$. The Hilbert space representation of 
$\cZ ^1 {}_h(\tau)= \Tr_{\H _h} q^{L_0} \bar q^{\tilde L_0}$ 
may be viewed as a spectral density function for the 
Hilbert space $\H_h$, which yields the conformal spectrum, including the
multiplicities for all states in $\H_h$. This essentially determines $\H_h$.

\item
The further application of modular transformations to $\cZ ^1 {}_h$ yields 
$\cZ ^{h^n}{}_h$, from which the action of the operators $h^n$ on $\H_h$
may be deduced.

\item
Combining the knowledge of the action of operators $h^n$ on $\H_h$
with the properties of the symmetric operators
in the associated symmetric orbifold theories has allowed us, 
in all cases considered here, to construct also the action 
of elements $g$ on $\H_h$, even when $g$ is not a power of $h$.
This last step permits us to calculate the remaining partition traces 
$\cZ ^g {}_h$.

\end{enumerate}

\medskip

The twisted Hilbert space ${\cal H}_h$ which emerges from the
above construction is of the following form. Let $G$ be the 
asymmetric orbifold group whose elements consist of pairs
$h=(h_L;h_R)$, where $h_L$ and $h_R$ act respectively on 
the left and the right sectors. The left and right elements themselves 
span groups $G_L$ and $G_R$ and $G$ may be viewed as
a subgroup of the product $G_L \times G_R$. The starting point
is the two {\sl symmetric orbifold theories} with symmetric
orbifold groups $G_L$ and $G_R$ consisting of pairs
$(h_L;h_L)$ and $(h_R;h_R)$ respectively.
Let ${\cal H}_{h_L} (p_L)$ and
${\cal H}_{h_R} (p_R)$ be the Hilbert spaces of the sectors
twisted by $(h_L,h_L)$ and $(h_R,h_R)$ in the
symmetric theories. Here, we denote by $p_L \in M_L$ 
and $p_R \in M_R$ any additional labels of the blocks,
such as for example the internal loop momenta.
$M_L$ and $M_R$ may be viewed as the degeneracies of the 
ground states in the symmetric theories. 
Then the Hilbert space of the $h$-twisted
sector in the asymmetric theory is of the form
\be
{\cal H}_h=\bigoplus_{(p_L,p_R)\in {\cal I}}\ \bigg (
{\cal H}_{h_L} (p_L) \,\otimes\,
{\cal H}_{h_R} (p_R) \biggr )
\ee
Here, ${\cal I}$ is a pairing or {\sl  selection rule}, and we keep
only chiral oscillators in ${\cal H}_{h_L} (p_L)$ and
${\cal H}_{h_R} (p_R)$. The set ${\cal I}$ is usually 
strictly smaller than the set of all $M_L \times M_R$ possible 
choices for $(p_L,p_R)$. It is determined by modular transformations
from the asymmetric traces in the untwisted sector. In \cite{nsv}, 
it was proposed to obtain the partition function
of the asymmetric theory by taking square roots of suitably
extended symmetric theories. Our prescription is
a significant departure from this, since it avoids square roots altogether and 
replaces them by selection rules after a suitable chiral splitting.

\medskip
 
The remainder of this paper is organized as follows. 
In section 2, a brief summary is given of the circle theory. 
In section 3, orbifolds generated by chiral shifts only are constructed. 
In section 4, orbifolds with chiral reflections only are investigated
and subtle issues of modular invariance are addressed and solved. 
In section 5, the results of \S 3 and \S 4 are combined and orbifolds 
with both chiral shifts and reflections are solved, and the effects of 
adding worldsheet fermions are included.
In section 6,  general rules for the partition function in higher genus
are formulated in terms of chiral splitting and a summation over 
pairings of chiral blocks.
Finally, in appendix  \S A,  the construction 
of symmetric orbifolds is briefly reviewed; in \S B, the equivalence is
demonstrated for symmetric orbifolds between orbifolding the full
line by the full orbifold group and orbifolding the circle by the point group;
and in \S C,  useful $\tet$-function identities are collected.

\section{The circle theory}
\setcounter{equation}{0}

We begin by recalling some basic facts about the $S^1$ theory, mainly for 
normalizations and conventions. The torus $\Sigma$ with modulus 
$\tau=\tau_1+i\tau_2$, $\tau_2>0$, is parametrized by $z$, 
with the identifications $z\sim z+1$ and $z\sim z+\tau$. It is equipped with
the metric $ds^2=\tau_2^{-1}dzd\bar z$. 
The action $S[x]$ for a theory of a single bosonic scalar field 
$x$ is normalized to be
\be
S [x]={1\over 4\pi \lbar ^2}\int_\Sigma d^2z  \p\,x \bar\p\,x,
\ee
where $\lbar$ is the string scale.\footnote{The parameter $\lbar$ is related 
to the Regge slope parameter $\alpha '$ by $2\lbar ^2=\alpha'$. 
Customary conventions used in the literature are as follows. 
In \cite{polchinski}, the Regge slope parameter $\alpha'$ is exhibited 
explicitly; 
in \cite{dp88} and \cite{dms}, one sets $\lbar=1$;  in \cite{g88}, 
one sets $\lbar=1/2$ instead.} 
In the $S^1$ theory, the field $x(z)$ takes values in a circle of 
radius $R$, and we have the identification $x\sim x+2\pi R$. 
The evaluation of the functional integral over all instanton 
sectors defined by the boundary conditions $x(z+1)=x(z)+2\pi m_1R$ 
and $x(z+\tau)=x(z)+2\pi m_2R$ gives the well-known partition function
\be
\label{real}
Z_{S^1_R} (\tau)
=
{R\over \lbar \sqrt{2\tau_2}|\eta(\tau)|^2}
\sum_{m_1,m_2\in {\bf Z}}{\rm exp}
\bigg\{-{\pi R^2\over 2\lbar ^2\tau_2}|m_1\tau-m_2|^2\bigg\}
\ee
Here, the argument of the exponential is the action of the corresponding 
instanton solution, and the prefactor combines the contributions of the 
quantum fluctuations and the zero mode integration. 
The partition function $Z_{S^1_R}$ can be recast in Hamiltonian 
language by a Poisson resummation in $m_2$
\be
\label{hamiltonian}
Z_{S^1_R} (\tau)
={1\over|\eta(\tau )|^2}
\sum_{m_1,m_2\in {\bf Z}}q^{{1\over 2}p_L^2}\bar q^{{1\over 2}p_R^2},
\hskip 1in
q\equiv e^{2\pi i\tau},
\ee
where the left and the right (dimensionless) momenta $p_L$ and $p_R$ are given 
by
\be
p_L={\lbar \over R}m_2-{R\over 2\lbar }m_1,
\qquad \quad
p_R={\lbar \over R}m_2+{R\over 2\lbar }m_1,
\qquad m_1,m_2 \in \bZ
\ee
In particular $(p_L,p_R)$ belongs to an even Lorentzian lattice 
$p_L^2-p_R^2=-2m_1m_2\in 2{\bf Z}$. 
(Note that the dimensionful momenta are given by $p_L/2\lbar$ 
and $p_R/2\lbar$, so that the total dimensionful momentum is 
$(p_L+p_R)/2 \lbar = m_2/R$, as expected.)

\medskip

Henceforth, we consider the $S^1$ theory at the self-dual radius 
$R^2=2\lbar ^2$. The left and right momenta simplify to
\be
\label{pn}
\s2 p_L=m_2-m_1=n_L,
\qquad
\s2 p_R=m_2+m_1=n_R,
\ee
with $n_L,n_R \in \bZ$ and $n_L+n_R\in 2{\bf Z}$. The momentum summations 
may be carried out in terms of $\tet$-functions. At the self-dual radius,
it is convenient to cast the $\tet$-functions results in terms of the 
$\zeta$-function, defined by
\bea
\label{zetadef}
\zeta [\alpha | \beta ] (\tau) 
\equiv 
{1 \over \eta (\tau)} \sum _{n\in \bZ} q^{ (n+\alpha )^2} e^{  2 \pi i n \beta}
= {\tet [\alpha |\beta] (0,2 \tau) \over \eta (\tau)} e^{- 2 \pi i \alpha \beta}
\eea
Various transformation properties of its arguments are listed in
Appendix C. One finds,
\bea
Z_{S^1_{R=\sqrt{2} \lbar}} (\tau) = 
|\zeta [0|0] (\tau)|^2 + |\zeta [\half |0] (\tau)|^2
\eea
where the first term arises from the contributions with $n_L,n_R\in 2 \bZ$
and the second term from contributions with $n_L,n_R\in 2\bZ+1$.

\medskip

The Hilbert space $\H$ 
of the circle theory can be built from chiral sectors as follows
\be
\label{untwistedsector}
\H
=
\bigoplus_{n_L+n_R\in 2{\bf Z}}
\biggl (
\biggl \{ \bigoplus_{n_j=1}^{\infty} \  x_{-n_1}\cdots x_{-n_k}|n_L\>_L
\biggr \} 
\bigotimes
\biggl \{ \bigoplus_{n_{\tilde j}=1}^{\infty} \  \tilde x_{-n_1}\cdots \tilde 
x_{-n_{\tilde k}}|n_R\>_R \biggr \} \biggr )
\ee     
Here, the states $|n_L\>_L, |n_R\>_R$ are momentum eigenstates, 
characterized by their quantum numbers $n_L,n_R$ as defined in (\ref{pn}),
and $x_{n_j}$, $\tilde x_{n_{\tilde j}}$ are oscillators
for each chiral sector. Then $Z_{S^1_R}$ becomes
\be
Z_{S^1_R} (\tau) = \Tr_\H ( q^{L_0}\bar q^{\tilde L_0})
\ee
with the following definitions of the Virasoro generators,
\bea
L_0 = -{1\over 24}+{1\over 2}p_L^2+\sum_{n=1}^{\infty}nx_{-n}x_n
\qquad \quad
\tilde L_0 = -{1\over 24}+{1\over 2}p_R^2+\sum_{n=1}^{\infty}n\tilde 
x_{-n}\tilde x_n
\eea
Finally, at any radius $R$, the circle theory has a chiral 
$\widehat {U(1)}_L \times \widehat {U(1)}_R$ symmetry, 
generated by the chiral currents $J^3 _L = \p x$ and $J^3 _R = \bar \p  x$. 
At the self-dual radius $R^2 = 2 \lbar ^2$,  the symmetry
is enhanced to $\widehat {SU(2)}_L \times \widehat {SU(2)}_R$, 
which  arises due to the existence of the extra currents 
$J^\pm _L = \exp \{ \pm i \s2 x_+\}$ and $J^\pm _R = \exp \{ \pm i \s2 x_-\}$, 
where $x_\pm$ denote the chiral parts of the field $x$. 
The latter symmetry will be exploited in section 4.2.

\section{Orbifolds defined by a chiral shift}
\setcounter{equation}{0}

The operator $s$ realizing a shift $x\to x+\pi R$ by a half-circumference 
commutes with the oscillators $x_n$ and  $\tilde x_n$ and acts on the 
momentum ground states by 
$s |n_L\>_L \otimes |n_R\>_R= e^{i\pi (p_L+p_R)R/(2\lbar 
)}|n_L\>_L\otimes|n_R\>_R$.
At the self-dual radius $R^2=2\lbar ^2$, this becomes simply 
\bea
s|n_L\>_L\otimes|n_R\>_R=e^{i\pi (n_L+n_R)/2}|n_L\>_L\otimes|n_R\>_R
\eea
This form naturally permits chiral splitting into actions of left and right 
chiral shift operators $s_L$ and $s_R$, which  also 
commute with the oscillators $x_n$ and $\tilde x_n$. Their action on the 
ground states is defined by 
\bea
\label{chirals}
s_L|n_L\>_L\otimes|n_R\>_R 
& = & 
e^{{1\over 2}i\pi n_L}|n_L\>_L\otimes|n_R\>_R
\no \\
s_R|n_L\>_L\otimes|n_R\>_R
&= &
e^{{1\over 2}i\pi n_R}|n_L\>_L \otimes|n_R\>_R
\eea
Clearly these definitions reproduce $s=s_Ls_R$.  Note  that each operator $s_L$ 
and $s_R$ has order 4 while their product $s$ has order 2 in view of the fact 
that $n_L+n_R\in 2 \bZ$. 
In this section,  orbifolds generated by either
$s_R$ or $s_R^2$ will be considered.

\subsection{The $s_R$ models}

Recall that the partition function of the orbifold theory defined by a finite 
abelian group $G$ is given (\ref{ZG}),  where $|G|$ is the order of $G$, and 
$\H_h$ is the Hilbert space of the twisted sector defined by $h$. As is reviewed 
in Appendix A,
in the case of the symmetric shift $s=s_Ls_R$, the Hilbert space in the twisted 
sector corresponding to an element $s^{-a}$ is generated by oscillators from
the ground states $|n_L+{a\over 2}\>\otimes |n_R-{a\over 2}\>$.

\medskip

In the asymmetric orbifold theory generated by $s_R$,  the Hilbert 
space of the  sector twisted by  the element $(s_R)^{-a}$ is defined to be the 
space generated by applying the oscillators $x_n$ and
$\tilde x_{-\tilde n}$ to the ground states of the form 
\bea
|n_L \>_L \otimes |n_R - {a \over 2}\>_R
\qquad \qquad 
n_L + n_R \in 2 \bZ
\eea
The untwisted sector may simply be viewed as the sector twisted by $(s_R)^{-a}$ 
for $a=0$, an observation that permits us to treat all sectors at once. This 
construction is consistent with the fact that $s_R$ is of order 4, since a shift
$a \to a+4$ may be compensated by relabeling the states by $n_R \to n_R +2$,
a transformation that preserves the condition $n_L+n_R\in 2 \bZ$.
The action of $(s_R)^b$ on a state in the sector twisted by 
$(s_R)^{-a}$ is given by
\bea
(s_R) ^b ~  |n_L \>_L \otimes |n_R - {a \over 2}\>_R
=
\exp \left \{ {i \pi b \over 2 } (n_R - {a \over 2}) \right \}
|n_L \>_L \otimes |n_R - {a \over 2}\>_R
\eea
with, as always, the constraint that $n_L + n_R \in 2 \bZ$.

\medskip

The partition traces for a single scalar field are easily calculated. The 
oscillator contributions are all equal to those of the
untwisted sector and result in the familiar $\eta$-function factors.
The momentum dependence may be read off from the momentum assignments
of the states. Abbreviating $\cZ^{g}{}_{h}=\cZ^b{}_a$ when $h=(s_R)^{-a}$ and 
$g=(s_R)^b$, the partition traces become,
\be
\label{defZ}
\cZ ^b {}_a (\tau) 
=  
{1 \over |\eta (\tau)|^2}
\sum _{n_L + n_R \in 2 \bZ} e^{ ib { \pi  \over 2} (n_R - {a \over 2})}
~ q^{{1 \over 4} n_L^2} ~ \bar q ^{{1 \over 4} (n_R - {a \over 2})^2}
\ee

The sums over $n_L$ and $n_R$ are conveniently expressed  in terms of the 
$\zeta$-function  defined in (\ref{zetadef}).
The constraint $n_L+n_R \in 2 \bZ$ in (\ref{defZ}) may be solved explicitly and 
gives rise to two parts : the first in which $n_L$ and $n_R$ are independent
even integers (first term on rhs) and the second in which they are 
independent odd integers (second term on rhs),
\bea
\label{parttrace}
\cZ ^b {}_a (\tau) & = &
{1 \over |\eta (\tau)|^2} \left ( 
\tet [0|0](0,2 \tau) \overline{\tet [{a \over 4}|{b \over 2}](0, 2 \tau)} +
\tet [\half |0](0,2 \tau) \overline{\tet [{a \over 4}+\half |{b \over 2}](0, 2 
\tau)}
\right )
\no \\
& = &
 e^{- {i \pi ab \over 4} }
\left ( \zeta [0|0](\tau) ~ \overline{\zeta [{a \over 4} | {b \over 2}](\tau)}
+ 
e^{- ib { \pi \over 2} } \zeta [\half |0](\tau) ~
\overline{\zeta [{a \over 4} +\half | {b \over 2}](\tau)} \right )
\eea  
From the partition traces of the $s_R$-twisted theory of a single boson
it is straightforward to obtain the partition traces of a theory of $d$ bosons
taking values in a square torus of all self-dual radii and where $s_R$  acts
as a right chiral shift simultaneously on all directions. As the $d$ bosons
in this model are independent of one another, the corresponding partition traces 
are simply given by $(\cZ^b {}_a)^d$.

\subsubsection{Interpretation in terms of chiral blocks constructions}

With the extension to higher loops in mind, it is useful to
interpret the previous formulas in the following manner.
We would like to view the traces $\cZ^b{}_a(\tau)$ of the
$s_L$ theory as built from the chiral blocks of the symmetric
$s=(s_L,s_R)$ theory and the untwisted theory, with a suitable
pairing. Now the partition function for the symmetric $s$ theory
at the self-radius $R^2=2\lbar^2$ is 
simply given by the partition function for the circle
theory at $R^2=\lbar^2/2$ and hence given by \cite{DVV}
\be
Z_{R^2=\lbar^2/2}^{\rm circle}(\tau) =\half \sum_{\epsilon,\delta} 
\cZ^{\delta}{}_\epsilon(\tau)
\ee
where $[\epsilon|\delta]$ runs over all half-characteristics,
and the ``trace" $\cZ^{\delta}{}_\epsilon(\tau)$ is
given by
\be
\cZ^{\delta}{}_\epsilon(\tau)
= {1 \over |\eta (\tau)|^2}
\sum_{\gamma\in \{0,\half\} }\big|\tet[\gamma+{1\over
2}\epsilon|\delta](0,2\tau)\big|^2
\ee
On the other hand, the partition function of the self-dual circle
model in arbitrary genus $h$ is given by
\be
Z_{R^2=2\lbar^2}^{circle} (\tau)
= {1 \over |\eta (\tau)|^2}
\sum_{\gamma\in \{ 0,\half\} ^h}
\big|\tet[\gamma|0](0,2\tau)\big|^2
\ee
where $\tau$ is now viewed as the period matrix and 
$\eta (\tau)$ as the chiral boson partition function at genus $h$. 
The formula (\ref{parttrace}) suggests that,
for fixed $b,a\in {\bf Z}_4=\{0,1,2,3\}$,
the chiral blocks of
the $s$ theory and the untwisted theory can be defined
respectively as
\bea
\label{chiralblocks}
Z_{R^2=\lbar^2/2}^{\rm s-theory}
& \qquad &
{1\over \eta(\tau)} \tet[\gamma+{1\over 4}a|\beta](0,2\tau)
\no \\
Z_{R^2=\lbar^2/2}^{\rm circle}:
& \qquad &
{1\over \eta(\tau)} \tet[\gamma|0](0,2\tau)
\eea
where $\gamma\in\{0,\half\}$ and we have expressed $b$ as
\be
\label{b}
b=2(\beta+c), \ \beta\in\{0,\half\},
\ c\in\{0,1\}.
\ee
With this choice of phases for the chiral blocks
arising from the symmetric theories, the blocks 
$\cZ^b{}_a(\tau)$
of the asymmetric $s_L$ theory arise by pairing them
with the matrix 
\bea
K_{\gamma\bar\gamma}
=\delta_{\gamma\bar\gamma}e^{2\pi ic\gamma}
\eea
so that we indeed recover the earlier partition traces, 
\be
\cZ^b{}_a(\tau)
=
{1\over|\eta(\tau)|^2}\sum_{\gamma\in \{ 0,\half \}}
e^{2\pi i c\gamma}
\tet[\gamma+{1\over
4}a|\beta](0,2\tau)\overline{\tet[\gamma|0](0,2\tau)}.
\ee
The full partition function for a theory of $d$ bosons is then given by
\be
Z(\tau)
=
{1\over 4}\sum_{a,b\in {\bf Z}_4}
\bigg\{{1\over|\eta(\tau)|^2}\sum_{\gamma\in \{0,\half \} }
e^{2\pi i c\gamma}
\tet[\gamma+{1\over
4}a|\beta](0,2\tau)\overline{\tet[\gamma|0](0,2\tau)}\bigg\}^d.
\ee
The partition function for the $s_R$ theory for $d$ bosons is
of course obtained by complex conjugation.

\subsubsection{Modular invariance of the $s_R$ model for $d\in 16{\bf N}$ 
bosons}

The modular covariance properties (\ref{modularcovariance}) of the partition 
traces follow  from the modular transformation laws (\ref{zeta1}) for the 
functions $\zeta[\alpha|\beta](\tau)$   and the relations  (\ref{zeta0}) and 
(\ref{zeta2}). 
One finds, 
\bea
\cZ ^b {}_a (\tau +1) & = & e^{i \pi a^2/8} \cZ ^{a+b} {}_a (\tau)
\no \\
\cZ ^b {}_a (-1/\tau ) & = & e^{- i \pi ab/4} \cZ ^a {}_{-b} (\tau)
\eea
Note that the exponent $a^2/16$ is as expected in the $\tau \to \tau+1$ 
transformation law because the ground state in the sector twisted by 
$(s_R)^{\pm a}$ has conformal dimension $a^2/16$ mod 1. As a result, 
the operator $s_R$ in the twisted sector actually has order 16 with $s_R^8=-1$.
To properly restore the order of $s_R$ to be 4 as it was in the untwisted 
sector,
the dimension of the torus (see the last paragraph of the preceding section) 
must be divisible by 4.

\medskip

More critically,  the orbifold model for a single boson does not satisfy the 
modular 
covariance requirement of (\ref{modularcovariance}). Partition traces with full 
modular invariance are obtained only when the dimension of the torus satisfies
$d\in 16{\bf N}$.

\subsubsection{Asymmetry of the $s_R$ partition function for $d=16$ bosons}

Often, an orbifold constructed from the action of an asymmetric orbifold group 
still possesses a symmetric partition function, i.e. $Z(\tau) = 
\overline{Z(\tau)}$. It will be verified below that the orbifold theory 
constructed from $16$ bosons and the asymmetric group action of $s_R$ is in fact 
asymmetric and 
satisfies $Z(\tau) \not= \overline{Z(\tau)}$. The starting point is the 
partition function defined by the partition traces of (\ref{parttrace}),
\bea
\label{partitions_R}
Z(\tau) 
=
{1 \over 4} \sum _{a,b=0,1,2,3} (\cZ^b {}_a (\tau) )^{16}
\eea
The most convenient study of the asymmetry issue is
in terms of $\tet$-functions of modulus $\tau$. To this end, $Z$
is first expressed in terms of $\tet$-functions of modulus $2 \tau$,
which are then converted into $\tet$-functions with modulus $\tau$
by using the doubling formulas. To simplify this calculation, note that 
a  shift $b\to b+2$ produces a relative sign between the two terms 
in the parenthesis of (\ref{parttrace}). It is convenient to restrict the range 
of $b=0,1$ and  isolate the summation over the shifts $b \to b+2$ using a new 
variable $c=0,1$. One obtains,
\bea
\left ( \cZ ^{b+2c} {}_a (\tau) \right )^2 =
{ 1 \over |\eta (\tau)|^4}
\left ( \tet [0|0] ~ \overline{\tet [ {a \over 4} | {b \over 2} ]}
+
(-)^c ~ \tet [\half |0] ~ \overline{\tet [\half +{a \over 4} |
{b \over 2} ]} \right )^2 (0,2\tau)
\eea
the partition function may be expressed as follows,
To compute the above partition traces in terms of $\tet$-functions with argument 
$\tau$, $\tet_i (0,\tau)$, with $i=2,3,4$,
we use the doubling formula for $\tet$-functions given in Appendix C.
One finds (here $a,c=0,1$)
 \bea 
\left ( \cZ ^{2c} {}_{2a} (\tau) \right )^2 
& = & 
{1 \over 2 |\eta|^4}
 \left ( |\tet _3|^4 +(-)^a |\tet
_4|^4 + (-)^c |\tet_2|^4 \right )
\no \\
\left ( \cZ ^{1+2c} {}_{2a} (\tau) \right )^2
 & = &
{1 \over 2 |\eta|^4} \left (\tet _3^2 + (-)^a \tet_4^2 \right ) \bar \tet _3 
\bar \tet _4
\no \\
\left ( \cZ ^{2c} {}_1 (\tau) \right )^2= \left ( \cZ ^{2c} {}_3 (\tau) \right 
)^2 
 & = & 
 {1 \over 2 |\eta|^4} \left (\tet _3^2 + (-)^c \tet _2^2 \right )\bar \tet_2 
\bar \tet_3
\no \\
\left ( \cZ ^{1+2c} {}_1 (\tau) \right )^2 = \left ( \cZ ^{1+2c} {}_3 (\tau) 
\right )^2 
 & = &
{1 \over 2 |\eta|^4} \left (-i \tet _4^2 + (-)^c \tet _2^2 \right )\bar \tet_2 
\bar \tet_4
\eea
Combining all contributions, one obtains
\bea
Z & = & { 1 \over 2^{10} |\eta (\tau)|^{32}}
\biggl \{
(|\tet_3|^4 + |\tet_4|^4 +|\tet_2|^4)^8 +
(|\tet_3|^4 + |\tet_4|^4 -|\tet_2|^4)^8
\no \\ && \hskip .8in
 +
(|\tet_3|^4 - |\tet_4|^4 +|\tet_2|^4)^8 +
(|\tet_3|^4 - |\tet_4|^4 -|\tet_2|^4)^8
\no \\ && \hskip .8in
+ 2 [(\tet _3^2 + \tet_4^2)^8 + (\tet _3^2 - \tet_4^2)^8]
  (\bar \tet _3 \bar \tet_4)^8
\no \\ && \hskip .8in
+ 2 [(\tet _3^2 + \tet_2^2)^8 + (\tet _3^2 - \tet_2^2)^8]
  (\bar \tet _3 \bar \tet_2)^8
\no \\ && \hskip .8in
+ 2 [(\tet _2^2 -i \tet_4^2)^8 + (\tet _2^2 +i \tet_4^2)^8]
  (\bar \tet _2 \bar \tet_4)^8 \biggr \}
\eea
Using (\ref{theta1}), it is straightforward to verify that 
$Z$ is modular invariant.

\medskip

 The partition function $Z$ for the 16-dimensional orbifold by $(1,s_R)$ is
asymmetric. To see this, it suffices to show that $Z- \bar Z \not= 0$.
The contribution of the first 4 terms in $Z$  is manifestly 
symmetric and cancels out of $Z- \bar Z$. The remaining terms may 
be simplified using the Jacobi identity $\tet _3^4 = \tet _2^4 + \tet _4^4$, and 
expressed as a function of a single homogeneous variable
$t \equiv \tet _4 ^4 / \tet _3^4$,
\bea
Z - \bar Z = {15 |\tet _3 |^{32} \over 2^7 |\eta (\tau)|^{32}}  (t-\bar
t) (1-t \bar t) (1 -t -\bar t) (t+\bar t - t\bar t)
\eea
Manifestly, this is not zero and the model has asymmetric conformal weights.

\subsection{The $s_R^2$ models}

The $s_R^2$ can be easily derived as a subsector of the
$s_R$ models, consisting of the blocks
$\cZ^b{}_a(\tau)$ with $a,b=0,2$.
The Hilbert space interpretation of all the blocks
is a direct consequence of the Hilbert space interpretation
provided for the $s_R$ theory. Because fewer blocks
appear in the $s_R^2$ theory, the requirement of
modular invariance block by block is less restrictive.
Explicitly,
\bea
\cZ^0{}_0(\tau)&=&|\zeta[0|0]|^2+|\zeta[{1\over 2}|0]|^2
\nonumber\\
\cZ^2{}_0(\tau)&=&|\zeta[0|0]|^2-|\zeta[{1\over 2}|0]|^2
\nonumber\\
\cZ^0{}_2(\tau)&=&\zeta[0|0]\overline{\zeta[{1\over 2}|0]}+
\zeta[{1\over 2}|0]\overline{\zeta[0|0]}
\nonumber\\
\cZ^2{}_2(\tau)&=&e^{-i\pi}(\zeta[0|0]\overline{\zeta[{1\over 2}|0]}-
\zeta[{1\over 2}|0]\overline{\zeta[0|0]})
\eea
Level matching requires the number $d$ of bosonic fields
to be a multiple of $4$, in which case we have modular
covariance of the partition traces. The resulting
partition function is
\bea
Z(\tau)
=
{1\over 2} \sum _{\sigma =\pm 1}
(|\zeta[0|0]|^2+ \sigma |\zeta[{1\over 2}|0]|^2)^d
+ {1\over 2} \sum _{\sigma =\pm 1}
(\zeta[0|0]\overline{\zeta[{1\over 2}|0]}+ \sigma
\zeta[{1\over 2}|0]\overline{\zeta[0|0]})^d
\eea
for $d\in 4{\bf N}$.
We observe that, although the orbifold group consists
of asymmetric elements, the resulting partition function
$Z$ turns out to be symmetric in this case.
For the minimal dimension of $d=4$, one may alternatively express 
the partition function in terms of $\tet$-function with modulus $\tau$.
The result is
\bea
Z (\tau) = { 1 \over 2 |\eta |^8} \left \{
|\tet _2 |^8 + |\tet _3 |^8 +|\tet _4 |^8  \right \}
\eea
It is easy to see that this partition function equals  the 
partition function for 8 decoupled Majorana fermions with
$SO(8)$ invariant action.

\section{Orbifolds defined by a chiral reflection}
\setcounter{equation}{0}

The symmetric reflection $r: x(z)\to -x(z)$ is realized as a unitary operator 
on the untwisted Hilbert space $\H$ by
$r|n_L\>_L\otimes |n_R\>_R= \pm |-n_L\>_L\otimes |-n_R\>_R$, and on the 
oscillators by $rx_{-n}=-x_{-n}r$, $r\tilde x_{-n}=-\tilde x_{-n}r$.
A chiral reflection $r_L$ should reflect $n_L$ while preserving $n_R$, while 
a chiral reflection $r_R$ should reflect $n_R$ while preserving $n_L$. Although
these actions of the chiral operators $r_L$ and $r_R$ are natural and unique,
the splitting $r=r_Lr_R$ generally leaves some phases undetermined.
Truly chiral operators will be obtained only if the phases associated with $r_L$
only depend upon $n_L$ (and the phases associated to $r_R$ only depend
on $n_R$). 
It may not always be possible to achieve this, in which case the splitting
of the operator $r = r_Lr_R$ may be viewed as {\sl anomalous}. 
Here, models will be sought which are anomaly free. Thus, the following 
action of $r_L$ and $r_R$ will be postulated,
\bea
\label{rLrR}
&&
r_L|n_L\>_L\otimes |n_R\>_R
=\rho _L(n_L)|-n_L\>_L\otimes |n_R\>_R
\no \\ &&
r_R|n_L\>_L\otimes |n_R\>_R
=\rho _R (n_R)|n_L\>_L\otimes |-n_R\>_R
\no \\ &&
r_Lx_{-n}=-x_{-n}r_L
\qquad
r_L\tilde x_{-n}=\tilde x_{-n}r_L
\no \\ &&
r_Rx_{-n}=x_{-n}r_R
\qquad
r_R\tilde x_{-n}=-\tilde x_{-n}r_R
\eea
where $\rho_L$ and $\rho_R$ are phases which may depend on $n_L$ and
$n_R$ respectively. 
The requirement that $r=r_Lr_R$ is a further constraint on the phases : 
$\rho_L(n_L) \rho_R(n_R) =\pm 1$ whenever $n_L+n_R \in 2 \bZ$. 
Furthermore,  note that the choice of $\rho_L(n_L)$ and $\rho_R(n_R)$ 
dictates the order of the chiral operators $r_L$ and $r_R$. 
It will be shown next that -- contrary to naive
expectation -- the order of the operators $r_L$ and $r_R$ 
cannot be 2, but must be larger.

\subsection{The order of the chiral operator $r_L$}

In order for the expression (\ref{ZG}) to correspond to a trace
of projections onto invariant subspaces, the operators $r_L$ and $r_R$ should
have the same order in all twisted sectors.
Therefore, it suffices to rule out the order $2$ in the untwisted sector. 
In this sector, the Hilbert space is well-known, and the partition traces for 
the  $r_L$ theory easily computed. 
The partition trace $\cZ^1{}_1$ is, of course, the same as $\cZ^0{}_0$  in 
(\ref{parttrace}), and $\cZ^{r_L}{}_1$ may be computed using the
definition of $r_L$ given above,
\bea
\label{ZrL1}
\cZ^1 {}_1 (\tau) & = & |\zeta [0|0]|^2(\tau) + |\zeta [\half| 0]|^2(\tau)
\no \\
\cZ^{r_L}{}_1(\tau) & = &
\rho _L (0)\zeta[0|{1\over 2}](\tau) \,\overline{\zeta[0|0]}(\tau)
\eea
In the second line above, the insertion of $r_L$ in the trace causes only the 
states generated from $n_L=0$ to contribute, and produces a factor $(-)^k$ 
when acting on the state 
$x_{-n_1}\cdots x_{-n_k}|n_L\>_L$. Note that
the traces in the untwisted sector only involve  $\rho_L(0)$.
Under $\tau \to \tau+1$, the partition trace $\cZ^{r_L}{}_1$ is invariant, as 
expected.

\medskip

It will be assumed that the partition traces satisfy modular covariance
(possibly when raised to some critical power $d$), and it will be shown that the 
assumption $r_L^2=1$ leads to a contradiction. 
Indeed, if $r_L^2=1$,  only the traces $\cZ^1{}_{r_L}$ and 
$\cZ^{r_L}{}_{r_L}$ in the sector twisted by $r_L$ remain to be determined. 
Modular covariance gives,
\bea
\label{ZrL2}
&&
\cZ^1{}_{r_L}(\tau)
=\cZ^{r_L}{}_1(-1/\tau)=
\rho_L (0)\,\zeta[{1\over 4}|0]\,
\biggl ( \overline{ \zeta[0|0]}+\overline{\zeta[{1\over 2}|0]  } \biggr ) (\tau)
\nonumber\\
&&
\cZ^{r_L^{-1}}{}_{r_L}(\tau)
=
\cZ^1{}_{r_L}(\tau+1)
=
\rho_L (0)\,e^{i\pi/8}\zeta[{1\over 4}|{1\over 2}]\,
\biggl ( \overline{\zeta[0|0]}+e^{-i\pi/2} \overline{ \zeta[{1\over 2}|0]} 
\biggr ) (\tau)
\eea 
Assuming that $r_L$ is of order two, ($r_L^2=1$ and thus $r_L^{-1}=r_L$) 
implies that,
$\cZ^1{}_{r_L}(\tau+2)=\rho_L (0)\,e^{i\pi/4}\,\zeta[{1\over 4}|0]
( \overline{\zeta[0|0]} - \overline{\zeta[{1\over 2}|0]})(\tau)$. 
But this result is clearly distinct from $\cZ^1{}_{r_L}(\tau)$, and hence 
violates
modular invariance and in particular the level 
matching condition of (\ref{levelmatching}), irrespective of the dimension $d$. 
Therefore, the order of
 $r_L$ cannot be 2. On the other hand, an analogous calculation yields
 $\cZ^1{}_{r_L}(\tau+4)= e^{i\pi/2} \cZ ^1 _{r_L} (\tau)$. Therefore, $r_L$ is 
 of order 4 provided that additionally the dimension $d$ is a multiple of 4 and 
 that $\rho _L(0)^4=1$.

\subsection{Chiral reflections of order $4$}

Henceforth,  it will be assumed that $r_L$ is of order $4$, and
that the dimension $d$ is a multiple of~$4$. In the preceding subsection, 
it was shown that these conditions are necessary  in order to have modular 
invariance and a  consistent Hilbert space formulation.
It was also necessary for the factor $\rho _L(0)$ to be a $4$-th root of unity.
Since all effects of $\rho_L(0)$ will disappears from all the partition traces 
in
all dimensions $d$ which are multiples of 4, we may simply set $\rho _L(0)=1$ 
without loss of generality. 
The traces $\cZ^1{}_1$ and $\cZ^{r_L}{}_1$ are unchanged from 
(\ref{ZrL1}) but are now considered with $\rho _L(0)=1$. Using modular 
covariance and charge conjugation symmetry, one has the following traces,
\bea
\label{ZrL3}
&&
\cZ^{r_L}{}_1(\tau) = \cZ^{r_L^3}{}_1(\tau) =
\zeta[0|{1\over 2}](\tau) \,\overline{\zeta[0|0]}(\tau)
\\
&&
\cZ^1{}_{r_L}(\tau) = \cZ^1{}_{r_L^3}(\tau)
=\cZ^{r_L}{}_1(-1/\tau)=
\zeta[{1\over 4}|0]\,
\biggl ( \overline{ \zeta[0|0]}+\overline{\zeta[{1\over 2}|0]  } \biggr ) (\tau)
\no \\
&&
\cZ^{r_L^{-b}}{}_{r_L}(\tau) = \cZ^{r_L^{b}}{}_{r_L^3}(\tau)
=
\cZ^1{}_{r_L}(\tau+b)
=
e^{i\pi b/8}\zeta[{1\over 4}|{b\over 2}]\,
\biggl ( \overline{\zeta[0|0]}+e^{-ib\pi/2} 
\overline{ \zeta[{1\over 2}|0]} \biggr ) (\tau)
\no \\
&&
\cZ^{r_L}{}_{r_L^2}(\tau) = \cZ^{r_L^3}{}_{r_L^2}(\tau)
=
\cZ^{r_L^2}{}_{r_L}(-1/\tau)
=
e^{i\pi /4}\zeta[0|\half]\,
\overline{ \zeta[{1\over 2}|0]}  (\tau)
\no
\eea 

Note that the partition traces in the untwisted sector and in the sector 
twisted by $r_L^2$ indeed yield different results.

\subsubsection{The sector twisted by $r_L$}

The next key step is to construct a Hilbert space $\H_{r_L}$ for the sector 
twisted by $r_L$ which can reproduce the traces $\cZ^1{}_{r_L}$ and 
$\cZ^{r_L}{}_{r_L}$.
Here, we encounter the key difficulty of how to determine the pairing between 
left and right movers in the twisted sector. Specifically, the chiral operator 
$r_L$ produces two twisted ground states, which may both be taken to be 
eigenstates, $r_L|\pm\>_L\sim |\pm\>_L$. Therefore,  the Hilbert space in the 
sector twisted by $r_L$ should be a {\it subspace} of the full direct 
product space of left and right movers,
\bea
\label{tildecHrL}
\tilde\H_{r_L}
&=&
\big\{x_{-n_1+{1\over 2}}\cdots x_{-n_k+{1\over 2}}|+\>_L
\otimes
\tilde x_{-n_1}\cdots\tilde x_{-n_{\tilde k}}|n_R\>_R, \ n_R\in{\bf Z}\big\} 
\nonumber\\
&&
\quad
\oplus\big\{x_{-n_1+{1\over 2}}\cdots x_{-n_k+{1\over 2}}|-\>_L
\otimes
\tilde x_{-n_1}\cdots\tilde x_{-n_{\tilde k}}|n_R\>_R,\
n_R\in\bZ\big\}
\eea
To determine which subspace $\H_{r_L}$ is the appropriate choice and how $r_L$ 
should act on this space,  use will be made of the partition traces 
$\cZ^1{}_{r_L}$ and $\cZ^{r_L}{}_{r_L}$, which were obtained earlier by modular 
covariance. 

\medskip

The partition trace $\cZ^1{}_{r_L}$ plays the role of a {\sl spectral density}, 
as each state contributes a factor $q^{L_0} \bar q^{\tilde L_0}$ times its 
multiplicity factor.
The spectral density (for a theory of bosons only) suffers no accidental
cancellations and may be used as an accurate guide for how left and right
chiralities are to be paired against one another. The expansion of 
$\cZ^1{}_{r_L}$ 
in powers of $q$ and $\bar q$ suggests defining $\H_{r_L}$ by the following 
pairing,
\bea
\label{cHrL}
\H_{r_L}
&=&
\big\{x_{-n_1+{1\over 2}}\cdots x_{-n_k+{1\over 2}}|+\>_L
\otimes
\tilde x_{-n_1}\cdots\tilde x_{-n_{\tilde k}}|n_R\>_R, \ n_R\in\,2{\bf Z}\big\} 
\nonumber\\
&&
\quad
\oplus\big\{x_{-n_1+{1\over 2}}\cdots x_{-n_k+{1\over 2}}|-\>_L
\otimes
\tilde x_{-n_1}\cdots\tilde x_{-n_{\tilde k}}|n_R\>_R,\
n_R\in\,2\bZ+1\big\}
\eea
The operator theoretic trace $\Tr_{\H_{r_L}}(q^{L_0}\bar q^{\bar L_0})$ agrees 
then with $\cZ^1{}_{r_L}(\tau)$, up to a factor of $e^{i\pi/8}$. 
Taking the dimension $d$ to be a multiple of $16$,  complete agreement is 
obtained. 

\medskip

More generally, the partition traces $\cZ^{r_L^b}{}_{r_L}$, with $b \in \bZ$ 
play the role of a spectral density weighed by the eigenvalues of the operator 
$r_L^b$ on the subspace $\H_{r_L}$,
as each state contributes a factor $r_L^b q^{L_0} \bar q^{\tilde L_0}$ times its 
multiplicity factor. 
Clearly, when $b$ is not a multiple of 4, this quantity may suffer 
from accidental cancellations
between states with identical conformal weight but different 
$r_L^b$ eigenvalue. Therefore, this partition trace will not, in general,
determine the action of $r_L^b$ is  a unique manner, and one will have to 
restrict to searching for internally consistent choices. Taking the operator 
$r_L$ 
to have order 4 on $\H_{r_L}$, one has the following natural assignment,
\bea
\label{rLtwisted}
r_L|+\>_L \otimes |n_R\>_R &=&|+\>_L \otimes |n_R\>_R,
\hskip 1in n_R\in 2\bZ
\nonumber\\
\ \
r_L|-\>_L \otimes |n_R\>_R &=& e^{-{1\over 2}i\pi}|-\>_L\otimes |n_R\>_R,
\hskip .62in  n_R\in 2\bZ+1
\eea
and the usual commutation relations with the oscillators.
With this,  the evaluation of all traces in the sector twisted by $r_L$ may be 
completed, and one finds for $b \in \bZ$,
\bea
\label{rLtwisted1}
\cZ^{r_L^b}{}_{r_L} (\tau)
=
 \zeta[{1\over 4}|{b \over 2}] \left ( \overline{\zeta[0|0]}+e^{i\pi b/2}
\overline{\zeta[{1\over 2}|0]} \right ) (\tau)
 \eea
For $b=0,1$, this formula indeed reproduces the result of (\ref{ZrL2}),
(which was obtained by modular transformations from partition traces in the 
untwisted sector) up to an overall factor of $e^{-i \pi b/8}$. Therefore, the 
Hilbert space formulation of the operator $r_L$ in the sector twisted by $r_L$,
given above, will be consistent with modular covariance only if the 
dimension $d$ is a multiple of 16.

\subsubsection{The sector twisted by $r_L^3$}

Charge conjugation symmetry, represented by the operator ${\cal C}= S^2$, 
ensures that the partition traces behave naturally under the reversal 
of orientation of the homology cycles, and implies the following relation on 
the partition traces,
$\cZ^g{}_h (\tau) = \pm \cZ^{g^{-1}}{}_{h^{-1}} (\tau)$. As the dimension  
is always assumed to be even, the $\pm$ factor is immaterial
and will be omitted.

\medskip

The charge conjugation relation may be used to investigate the partition
traces and the Hilbert space in the sector twisted by $r_L^3= r_L ^{-1}$.
Applying charge conjugation to (\ref{rLtwisted1}), it is readily derived that
for $b \in \bZ$,
\bea
\label{rL3twisted1}
\cZ^{r_L^b}{}_{r_L^3} (\tau)
=
 \zeta[{1\over 4}|{b \over 2}] \left ( \overline{\zeta[0|0]}+e^{- i\pi b/2}
\overline{\zeta[{1\over 2}|0]} \right ) (\tau)
\eea
Notice the difference in the relative phase factor between the 
two terms in the parenthesis above when $b$ is odd.
As a result of this phase difference the structure of the Hilbert space 
$\H_{r_L^3}$ is 
very close to that of $\H_{r_L}$ but not identical, as the operator
$r_L$ has a different action on both spaces. Therefore, extra care is 
needed in constructing also this space correctly. The two 
degenerate ground states
in the sector twisted by $r_L^3$ will be denoted by $|\pm \>_L ^\C$,
where the superscript $\C$ stands for charge conjugation.
The Hilbert space is  
\bea
\label{cHrL3}
\H_{r_L^3}
&=&
\big\{x_{-n_1+{1\over 2}}\cdots x_{-n_k+{1\over 2}}|+\>_L^\C
\otimes
\tilde x_{-n_1}\cdots\tilde x_{-n_{\tilde k}}|n_R\>_R, \ n_R\in\,2{\bf Z}\big\} 
\nonumber\\
&&
\quad
\oplus\big\{x_{-n_1+{1\over 2}}\cdots x_{-n_k+{1\over 2}}|-\>_L^\C
\otimes
\tilde x_{-n_1}\cdots\tilde x_{-n_{\tilde k}}|n_R\>_R,\
n_R\in\,2\bZ+1\big\}
\eea
and the analysis of the partition traces in (\ref{rL3twisted1})
leads to the following definition of the action of the operator
$r_L$ on $\H _{r_L^3}$,
\bea
\label{rL3twisted}
r_L|+\>_L^\C \otimes |n_R\>_R &=&|+\>_L^\C \otimes |n_R\>_R,
\hskip 1in n_R\in 2\bZ
\nonumber\\
\ \
r_L|-\>_L^\C \otimes |n_R\>_R &=& e^{+{1\over 2}i\pi}|-\>_L^\C\otimes |n_R\>_R,
\hskip .62in  n_R\in 2\bZ+1
\eea
and the usual commutation relations with the oscillators.
With these assignments, all partition traces in $\H_{r_L^3}$ are consistent
with modular covariance and the values of the remaining partition traces
$\cZ^{r_L^3}{}_1=\cZ^{r_L}{}_1$.

\subsubsection{The sector twisted by $r_L^2$}

It remains to analyze the sector twisted by $r_L^2$. Since $r_L$ is 
not of order 2 but rather 4, the Hilbert space $\H_{r_L^2}$ in the sector
twisted by $r_L^2$ is expected to be distinct from the untwisted Hilbert space. 
The starting point is
the partition trace with $r_L^2$ inserted in the untwisted sector.
The eigenvalues of $r_L^2$ in the untwisted sector are 
deduced from (\ref{rLrR}), 
\bea
r_L^2 |n_L\>_L \otimes |n_R\>_R = 
\rho _L (-n_L) \rho _L (n_L)  |n_L\>_L \otimes |n_R\>_R
\eea
Since $r_L^4=1$, the combination $\rho _L (-n_L) \rho _L (n_L)$
can take eigenvalues $\pm 1$ only. There are two natural 
assignments,
\bea
\left \{ \matrix{
{\rm I} &  & \rho _L (-n_L) \rho _L (n_L) = 1 
& {\rm for \ all \ } n_L \in \bZ \cr
{\rm II} &  & \rho _L (-n_L) \rho _L (n_L) = (-1)^{n_L} 
& {\rm for \ all \ } n_L \in \bZ \cr}
\right .
\eea
In the next subsection, it will be shown that these are not only natural
assignments, but in fact the only ones that are compatible with the
$\widehat{SU(2)}_L \times \widehat{SU(2)}_R$ symmetry of the theory.

\medskip

In case I, $r_L^2$ acts as the identity in the untwisted sector
and modular covariance requires $\cZ ^1 {}_1 = \cZ^{r_L^2}{}_1 = 
\cZ^{r_L^2}{}_{r_L^2} = \cZ^1{}_{r_L^2}$. The full partition function is 
therefore
\bea
\label{rLpartitionI}
Z_{\rm I}
&=&
\biggl  ( |\zeta[0|0]|^2+  | \zeta[{1\over 2}|0] |^2 \biggr )^d 
+\half \zeta[0|{1\over 2}]^d\overline{\zeta[0|0]^d}
+
\half \zeta[0|{1\over 2}]^d\overline{\zeta[{1\over 2}|0]^d}
\no \\ && +
{1 \over 2} \sum_{b=0}^3\zeta[{1\over 4}|{b\over 2}]^d
\biggl (
\overline{\zeta[0|0]+e^{\half i\pi b}\zeta[{1\over 2}|0]} \biggr ) ^d
\eea
This theory appears to have the undesirable property 
that the order of $r_L$ is 4 in the sector twisted by $r_L$ 
but that the order is only 2 in the untwisted sector.
The counting of the low lying states suggests that the theory is
inconsistent. 

\medskip

In case II, we have the following partition trace in the untwisted sector,
\bea
\cZ^{r_L^2}{}_1=|\zeta[0|0]|^2 -|\zeta[{1\over 2}|0]|^2
\eea
Modular covariance may be used to derive the remaining partition traces,
\bea
\cZ^1{}_{r_L^2} 
& = &
\zeta[0|0]\overline{\zeta[{1\over 2}|0]}
+
\overline{\zeta[0|0]}\zeta[{1\over 2}|0]
\no \\
\cZ^{r_L}{}_{r_L^2} = \cZ^{r_L^3}{}_{r_L^2}
&=&
\zeta[0|{1\over 2}] \overline{\zeta[{1\over 2}|0]}
\no \\
\cZ^{r_L^2}{}_{r_L^2}
&=&
\zeta[0|0]\overline{\zeta[{1\over 2}|0]}
-
\overline{\zeta[0|0]}\zeta[{1\over 2}|0]
\eea

Proceeding as in the other twisted sectors, the  Hilbert space   
$\H_{r_L^2}$ is obtained from $\cZ^1{}_{r_L^2}$,
\be
\H_{r_L^2}
=\{ x_{-n_1}\cdots x_{-n_k} \tilde x_{-n_1} \cdots \tilde x_{-n_{\tilde k}} 
|n_L \>_L \otimes |n_R-1 \>_R,
\ \ n_L+n_R\in 2\bZ\}
\ee
The trace over this Hilbert space of the operator $q^{L_0}\bar q^{\bar L_0}$
indeed reproduces $\cZ^1{}_{r_L^2}$. The  action of $r_L$ on
$\H_{r_L^2}$ is inferred from $\cZ^{r_L^b}{}_{r_L^2} $. The following 
phase assignment will be assumed, consistently with the fact that
the chiral parts of $\H_{r_L^2}$ are identical to the chiral parts of the 
untwisted sector,
\bea
r_L|n_L\>_L\otimes |n_R-1\>_R=
\rho _L(n_L) |-n_L\>_L\otimes |n_R-1\>_R
\eea
We continue to assume that $\rho _L(0)=1$ without loss of generality.
As only states with $n_L=0$ contribute to $\cZ^{r_L}{}_{r_L^2} = 
\cZ^{r_L^3}{}_{r_L^2}$, these partition traces are readily 
reproduced by the above assignments.
From the expression for $\cZ^{r_L^2}{}_{r_L^2}$, it is clear that
$r_L^2$ has eigenvalue $+1$ when $n_L$ is even, and $-1$
when $n_L$ is odd. Therefore, $\rho (n_L)$ must satisfy
\bea
\label{rhos}
\rho_L(-n_L) \rho (n_L)=(-1)^{n_L}
\eea
consistently with the fact that $r_L^4=1$. There is no canonical choice $\rho 
_L$ 
satisfying this relation; a convenient choice is $\rho _L(n_L) = \exp \{ i \pi 
|n_L|\}$,
which has the additional property that $\rho _L(n_L)^4=1$ for all $n_L\in \bZ$.

\medskip

Altogether, the $r_L$ theory with $d\in 16{\bf N}$ fields is  modular 
covariant, possesses a consistent Hilbert space interpretation and has the 
following full partition function
\bea
\label{rLpartition}
Z_{{\rm II}}
&=&
{1\over 4} \sum _{\sigma =\pm 1}
\biggl  ( |\zeta[0|0]|^2+ \sigma | \zeta[{1\over 2}|0] |^2 \biggr )^d 
+ {1\over 4} \sum _{\sigma =\pm 1}
\biggl ( \zeta[0|0]\overline{\zeta[{1\over 2}|0]}
+ \sigma  \overline{\zeta[0|0]} \zeta[{1\over 2}|0] \biggr )^d
\\
&&
+\half \zeta[0|{1\over 2}]^d\overline{\zeta[0|0]^d}
+
\half \zeta[0|{1\over 2}]^d\overline{\zeta[{1\over 2}|0]^d}
+
{1 \over 2} \sum_{b=0}^3\zeta[{1\over 4}|{b\over 2}]^d
\biggl (
\overline{\zeta[0|0]+e^{\half i\pi b}\zeta[{1\over 2}|0]} \biggr ) ^d
\qquad
\no
\eea
Term by term comparison reveals that this partition function 
is identical to (the complex conjugate of) the partition function
of the model twisted by $s_R$, given by combining (\ref{parttrace})
with (\ref{partitions_R}) for dimension $d=16$. The significance 
of this identification will be elucidated in the next section.

\subsection{The symmetry 
$\mathbf{\widehat{SU(2)}_L \times \widehat{SU(2)}_R}$}

At the self-dual radius, $R^2=2\lbar ^2$, the conformal field theory of a single 
boson possesses the symmetry 
$\widehat{SU(2)}_L \times \widehat{SU(2)}_R$. 
The generators of $\widehat{SU(2)}_L$ are the holomorphic currents 
\be
J^3_L  = \p x_+ \hskip 1in
\left \{ \matrix{
J^+_L = {1 \over \s2} (J^1 _L +i J^2_L) = e^{+ i \s2 x_+}
\cr
J^-_L = {1 \over \s2} (J^1 _L -i J^2_L) = e^{- i \s2 x_+}\cr } \right .
\ee
where $x_+$ denotes the left chiral part of the field $x$. Using the fact that
under $s_L$, one has the transformation $x_+ \to x_+ + \pi /\s2$, we are in a 
position to obtain the transformation laws of these generators under the chiral 
operators, (see e.g. \cite{dms}),
\bea
~[J^a_L, r_R] = 0 
\hskip .6in 
r_L J_L^1 r_L ^\dagger  = + J^1 _L 
\hskip .6in 
r_L J_L^{2,3} r_L ^\dagger  = - J^{2,3} _L
\no \\
~[J^a_L , s_R] = 0
\hskip .6in 
s_L J_L^3 s_L ^\dagger  = + J^3 _L 
\hskip .6in 
s_L J_L^{1,2} s_L ^\dagger  = - J^{1,2} _L
\eea
Thus, $r_L$ and $s_L$ both correspond to $SU(2)_L$ 
rotations by 180 degrees, $r_L$ about the 1-axis and $s_L$ about the 3-axis. 
This observation explains (1) why the orders of $r_L$ and $s_L$ must be 
equal -- to 4; (2) why the critical dimension of the asymmetric 
orbifold models constructed from them coincide -- and are multiples of 16; (3)
why their partition functions  coincide.

\medskip

The presence of this enhanced symmetry also has consequences
for the phase assignments $\rho _L(n_L)$ associated with $r_L$.
The operator $r_L^2$ corresponds to a 360 degree rotations and, 
as expected, commutes with the currents $J^a_L$ and must thus
assume constant eigenvalue throughout any irreducible representation
of $\widehat{SU(2)}_L$. In the untwisted sector, two representations
occur, one of spin 0 corresponding to $n_L$ even, and one of spin
1/2 corresponding to $n_L$ odd. The first representation of $r_L^2$
would be the trivial (identity) representation, but this was 
ruled out for being inconsistent with modular invariance. The
only other non-trivial representation is then, 
\bea
r_L^2 | n_L \> =  (-)^{ n_L} | n_L \>
& \qquad & 
r_L | n_R \> = | n_R \>
\no \\
r_R^2 | n_R \> =  (-)^{ n_R} | n_R \>
& \qquad & 
r_R | n_L \> = | n_L \>
\eea

Notice that these assignments are in complete analogy with the action of 
$s_{L,R}^2$ in the untwisted sector. For $s_{L,R}$, however, the phase
assignment of their squares dictated their own action. For $r_{L,R}$, 
the condition is weaker as the invariant generator $J^1 _L$ is not 
diagonal in the basis $|n_L\>_L$. It simply amounts to the relation
$ \rho _L (-n_L) \rho (n_L) = (-1)^{n_L}$, already derived in (\ref{rhos}).

\subsubsection{Interpretation in terms of chiral blocks
constructions}

As we did earlier for the asymmetric $s_L$ theory,
we can give an interpretation for the asymmetric
$r_L$ theory in terms of chiral blocks of the symmetric
$r=(r_L,r_R)$ theory.
Now the chiral operator $r_L$ is
of higher order than the symmetric operator $r=(r_L,r_R)$,
and at first sight, there seems to be no way of differentiating
between the conformal blocks of, say, $r_L^b$ and $r_L^{b+2}$.
But the one-loop case shows that the blocks
come with a specific assignment of phases, and it is these
phases which distinguish between the blocks of
$r_L^b$ and $r_L^{b+2}$.
More precisely, let $b,a\in {\bf Z}_4$ as before,
with $b$ expressed uniquely in terms of $\beta\in\{0,\half\}$
and $c\in\{0,1\}$ as in (\ref{b}). In \cite{DVV}, it was
shown that the orbifold of the circle theory at self dual radius
by $r=(r_L,r_R)$ coincides with the circle theory at half
the self-dual radius. Thus the chiral blocks of the $r=(r_L,r_R)$
theory and the chiral blocks of the untwisted theory are given
as before by (\ref{chiralblocks}). This time, our formulas
obtained earlier by Hilbert space methods show that they
are paired differently in order to give the
partial traces of the asymmetric $r_L$ theory. In fact, the
pairing matrix is $K_{\gamma\bar\gamma}
=\delta_{\gamma\bar\gamma}e^{2\pi i(c+2\beta)\gamma}$,
and the partial traces given by
\be
\cZ^b{}_a(\tau)
=
{1\over|\eta(\tau)|^2}
\sum_{\gamma\in\{0,\half\}}\,
e^{2i\pi (c+2\beta)\gamma}
\tet[\gamma+{1\over 4}a|\beta](0,2\tau)
\overline{\tet[\gamma|0](0,2\tau)},
\ee
Hence the partition function of the $r_L$ theory
for $d$ bosons is given by
\be
Z(\tau)
=
{1\over 4}\sum_{a,b\in {\bf Z}_4^g}
\bigg\{{1\over |\eta(\tau)|^2}\sum_{\gamma\in{1\over 2}{\bf Z}}
e^{2i\pi(c+2\beta)\gamma}
\tet[\gamma+{1\over
4}a|\beta](0,2\tau)
\overline{\tet[\gamma|0](0,2\tau)}\bigg\}^d.
\ee
As noted before, the partition functions of the $s_L$ and the
$r_L$ theories are the same, although the blocks
for the assignment $s_L^b$ and $s_L^a$ on the $B$ and $A$
cycles differ by phases from the blocks for the assignment
of $r_L^b$ and $r_L^a$ on the same cycles.

\section{Orbifolds defined by chiral shifts and reflections}
\setcounter{equation}{0}

The next step towards analyzing the models of Kachru-Kumar-Silverstein 
is the consideration of asymmetric orbifold models which involve
both chiral shifts $s_{L,R}$ and chiral reflections $r_{L,R}$. First, a detailed
study is presented of a purely bosonic model with only a single chiral 
generator $f \equiv (r_L,s_R)$; second, the model is extended to 
include two chiral generators $f$ and $g \equiv (s_L,r_R)$; third, fermions
are included as well and the full KKS model is analyzed.

\subsection{The $f=(r_L,s_R)$ model}

The proper definition of the chiral operator $f$ involves some of the 
same delicate issues that arose when defining $r_L$. First,
in the untwisted sector, $f$ will involve the phases $\rho _L (n_L)$,
\bea 
\label{phasef}
f |n_L \>_L \otimes |n_R \>_R 
& = & 
\rho _L(n_L) e^{i \pi n_R/2}  |-n_L \>_L \otimes |n_R \>_R
\no \\
f^2 |n_L \>_L \otimes |n_R \>_R 
& = & 
\rho _L(-n_L) \rho _L(n_L) (-1)^{n_R}  |n_L \>_L \otimes |n_R \>_R
\eea
Depending on whether  $\rho _L(-n_L) \rho _L (n_L)$ equals 1 or $(-1)^{n_L}$, 
the order of $r_L$ is 2 or 4, while the order of $f$ will be 4 or 2 respectively
(the latter since $n_L + n_R \in 2 \bZ$). 
Which of these choice (if any) leads to a consistent $f$-orbifold
is analyzed below.

\subsubsection{Ruling out the case of trivial phases $\rho _L (n_L)$}

It is first shown that the naive definition of $r_L$, 
for which $\rho _L(-n_L) \rho _L (n_L)=1$ for all $n_L \in \bZ$, 
and $f$ is of order 4, is incompatible with  the dual requirements of
modular covariance and a proper Hilbert space interpretation.
With this naive choice, one has  $f^2=(1,s_R^2)\not=1$. 
The $4$ partition traces in the untwisted sector are  readily evaluated to be
\bea
\label{traces}
\cZ^1{}_1
=|\zeta[0|0]|^2+|\zeta[{1\over 2}|0]|^2
&\qquad &
\cZ^f{}_1
= |\zeta[0|{1\over 2}]|^2
\no \\ 
\cZ^{f^2}{}_1
=
|\zeta[0|0]|^2-|\zeta[{1\over 2}|0]|^2
&\qquad &
\cZ^{f^3}{}_1
=
|\zeta[0|{1\over 2}]|^2
\eea
Note that the relative $-$ sign in $\cZ^{f^2}{}_1$ appears due to
the presence of $s_R^2$ in $f^2$.
Assuming modular covariance, all other partition traces may be derived
from (\ref{traces}); they are
\bea
\label{tracesf}
\cZ^1{}_f = \cZ^1{}_{f^3} = \cZ^{f^2}{}_f = \cZ^{f^2}{}_{f^3} 
& = &
2|\zeta[{1\over 4}|0]|^2
\no \\
\cZ^f{}_f = \cZ^{f}{}_{f^3} = \cZ^{f^3}{}_f = \cZ^{f^3}{}_{f^3}
& = &
2|\zeta[{1\over 4}|{1\over 2}]|^2
\eea
in the sectors twisted by $f$ and $f^3$, and  
\bea
\label{tracesf2}
\cZ^1{}_{f^2}
=
\zeta[0|0]\overline{\zeta[{1\over 2}|0]}
+
\overline{\zeta[0|0]}\zeta[{1\over 2}|0]
&\qquad &
\cZ^f{}_{f^2}(\tau)
=|\zeta[0|{1\over 2}]|^2
\no \\
\cZ^{f^2}{}_{f^2}
=
e^{-i\pi/2}(\zeta[0|0]\overline{\zeta[{1\over 2}|0]}
-
\overline{\zeta[0|0]}\zeta[{1\over 2}|0])
&\qquad &
\cZ^{f^3}{}_{f^2}(\tau)
=
|\zeta[0|{1\over 2}]|^2
\eea
in the sector twisted by $f^2$. Since  $f^2=(1,s_R^2)$, 
the Hilbert space of the sector twisted by $f^2$ may easily
be constructed and one has,
\be
\H_{f^2}
=
\{x_{-n_1}\cdots x_{-n_k}|n_L\>_L
\otimes
\tilde x_{-n_1}\cdots x_{-n_{\tilde k}}|n_R-1\>_R,
\ n_L+n_R\in 2\bZ\}
\ee
This allows an independent evaluation of $\cZ^f{}_{f^2}(\tau)$ 
as an operator trace; one finds,
\be
\Tr_{{\H}_{f^2}} \left (f q^{L_0}\bar q^{\bar L_0} \right )
=
-{1\over q^{1/24}\prod_{n=1}^{\infty}(1+q^n)}
\overline{\tet[{1\over 2}|{1\over 2}](0,\tau)\over \overline{\eta(\tau)}}
=0
\ee
This contradicts the earlier, non-vanishing, formula for $\cZ^f{}_{f^2}(\tau)$ 
obtained from modular covariance
\be
0\not=\cZ^f{}_1 \to \cZ^1{}_f \to \cZ^f{}_f \to \cZ^{f^2}{}_f
\to \cZ^f{}_{f^2} \not=0
\ee
Therefore, the assumption that the phase assignment of $r_L$ is 
trivial cannot lead to a consistent orbifold theory.

\subsubsection{Consistency of the case of non-trivial phases $\rho _L(n_L)$}

For the non-trivial phase assignment, 
$\rho _L(-n_L) \rho _L (n_L)=(-1)^{n_L}$,
one has  $f^2=1$ in the untwisted sector and the theory truncates.
The traces $\cZ^1{}_1$ and $\cZ^f{}_1$ in the untwisted sector do not depend
on the choice of $\rho _L(n_L)$ (having set $\rho _L(0)=1$  without loss of 
generality),  and are still given by their expressions in (\ref{traces}). Since 
the 
traces $\cZ^1{}_f$ and $\cZ^f{}_f$ follow by modular covariance, 
they are also given by their expressions in (\ref{tracesf}).

\medskip

The only issue that remains to be verified is that the traces in the sector 
twisted by $f$ can indeed be realized as partition traces in a twisted Hilbert
space $\H_f$, where $f$ is realized as an operator of order $2$. 
It is well-known \cite{g88,dms} that the chiral sector twisted by $r_L$ 
contains two degenerate ground states, which we shall denote by $|\pm \>_L$. 
Also, the chiral sector twisted by $s_R$ contains two sectors $|n_R -1/2\>_R$
with $n_R$ either even or odd.
Following the method of \S 4, the Hilbert space is found to be
\bea
\label{Hf}
\H_f
&=&
\bigg\{(x_{-n_1+\half}\cdots x_{-n_k+\half}|+\>_L
\otimes \tilde x_{-n_1}\cdots\tilde x_{-n_{\tilde k}}|n_R-\half\>_R,\ \ n_R\in 
2{\bf Z}\bigg\}
\\
&&
\quad 
\oplus
\bigg\{(x_{-n_1+\half}\cdots x_{-n_k+\half}|-\>_L
\otimes \tilde x_{-n_1}\cdots\tilde x_{-n_{\tilde k}}|n_R- \half\>,\ \ n_R\in 
2{\bf Z}+1\bigg\}
\nonumber
\eea
Note that the pairing between the $|\pm \>_L$ and $|n_R- 1/2>_R$ 
provides a highly non-trivial piece of information on how the left and 
right chiral Hilbert spaces need to be combined. The pairing indeed
implies that out of the 4 ground states arising from a full tensor
product of left and right, only 2 are to be retained in the asymmetric
orbifold theory. The same pairing issue arose already in the theory 
twisted by the symmetric operator $r=(r_L,r_R)$, where also only 2 states
out of a total of 4 should be retained.\footnote{ 
Although the issue is well-known in the symmetric theory \cite{g88}, 
it has nonetheless given rise to confusion when the 
proper pairing was not taken into account.}
It is readily verified that,
\bea
\cZ ^1 {}_f = \Tr_{{\H}_f}(q^{L_0}\bar q^{\bar L_0})
= 2\,|\zeta[{1\over 4}|0]|^2
\eea
as desired. To obtain $\cZ^f{}_f$ from $\H_f$ as well, it suffices to define 
$r_L$ on the momenta ground states in $\H_f$ as
\bea
\label{rLftwisted}
r_L|+\>_L \otimes |n_R - \half \>_R 
&=&
|+\>_L \otimes |n_R - \half \>_R 
\hskip 1in n_R \in 2 \bZ
\nonumber\\
r_L|-\>_L \otimes |n_R - \half \>_R 
&=&
-i\,|-\>_L \otimes |n_R - \half \>_R 
\hskip .8in n_R \in 2 \bZ +1
\eea  
and similarly for excited states. With this construction, 
$\Tr_{{\H}_f}(q^{L_0}\bar q^{\bar L_0})$ is indeed found to agree with the 
expression found earlier for $\cZ^f{}_f$ by modular covariance.

\medskip

Altogether, the theory obtained this way is modular covariant and admits a 
Hilbert space realization in terms of twisted sectors. 
Its partition function in dimension $d\in 4{\bf N}$ is given by
\be
\label{partf}
Z =
{1\over 2} \bigg\{
(|\zeta[0|0]|^2+|\zeta[{1\over 2}|0]|^2)^d
+
|\zeta[0|{1\over 2}]|^{2d}
+
2^d|\zeta[{1\over 4}|0]|^{2d}
+
2^d|\zeta[{1\over 4}|{1\over 2}]|^{2d}
\bigg\}
\ee
which coincides with that of the $d$-dimensional circle theory 
orbifolded by $r=(r_L,r_R)$.

\subsection{Orbifolds generated by $f$ and $g$}

Next, we consider the theory of $d$ bosonic scalar fields moded out by the group 
$G$ generated by the two elements $f=(r_L,s_R)$ and $g=(r_R,s_L)$.
This group is Abelian; indeed, in view of (\ref{phasef}), one has  in the 
untwisted sector
\bea
f |n_L \>_L \otimes |n_R \>_R 
& = & 
\rho _L(n_L) e^{i \pi n_R/2}  |-n_L \>_L \otimes |n_R \>_R
\no \\
g |n_L \>_L \otimes |n_R \>_R 
& = & 
\rho _R(n_R) e^{i \pi n_L/2}  |n_L \>_L \otimes |- n_R \>_R
\eea
and therefore
\bea
gf |n_L \>_L \otimes |n_R \>_R  = e^{i \pi (n_R-n_L)} fg |n_L \>_L \otimes |n_R 
\>_R 
\eea
In view of the fact that in the untwisted sector $n_L+n_R \in 2 \bZ$, the phase 
factor on the rhs is just 1, and one has $fg=gf$. Since also $f^2=g^2=1$, the 
orbifold group generated by $f,g$ is just the group $\bZ_2\times\bZ_2$. 
As usual, it taken for granted that the group composition laws remain the 
same in all sectors.

\subsubsection{The sectors twisted by $f$ and $g$}

The partition traces $\cZ^{f^n}{}_1$, (and by interchanging
left and right movers also $\cZ^{g^n}{}_1$), have already been determined in 
(\ref{traces}) and we have e.g.  $\cZ^f{}_1 = \cZ^g{}_1 = |\zeta [0|\half]|^2$.
By modular transformation, the partition traces are found in the twisted 
sectors, 
$\cZ ^1 {}_f = \cZ ^1 {}_g = 2 |\zeta [\quar |0]|^2$.
The Hilbert space $\H_f$ in the sector twisted by $f$, (and by interchanging
left and right movers also the Hilbert space $\H_g$ in the sector twisted by 
$g$) was given in (\ref{Hf}).

\medskip

The novel partition traces to be calculated in $\H_f$ are $\cZ ^g {}_f$ and 
$\cZ ^{fg} {}_f$. They belong to a modular orbit that {\sl does not include 
any partition trace evaluated in the untwisted sector}, and is thus not 
derivable 
from preceding results.
To evaluate $\cZ ^g {}_f$, it suffices to find $g$ on the ground states of $\H 
_f$,
\bea
\label{gaction}
g |\pm \>_L \otimes |n_R - \half \>_R 
& = &
\left ( s_L |\pm \>_L \right ) \otimes  ( r_R |n_R - \half \>_R  )
 \no \\
& \sim &
\left ( s_L |\pm \>_L \right ) \otimes  (  |- n_R + \half \>_R  ) 
\eea
Here, the $\sim$ sign has been used, because a phase factor
may arise when applying $r_R$ to $| \ \>_R$ states. The half-integer 
moding of the $| \ \>_R$ states alone guarantees that the operator 
$g$ has no diagonal matrix elements in $\H_f$, and must therefore
have vanishing trace,  $\cZ ^g {}_f=0$. Modular invariance 
then implies that all partition traces in the same orbit must vanish,
\bea
\label{fgvanish}
\cZ ^g {}_f = \cZ ^f {}_g = \cZ ^{fg}{}_f = \cZ ^f {}_{fg} =
\cZ ^{fg}{}_g = \cZ ^g {}_{fg} =0
\eea
In the sector $\H_f$, the vanishing of the partition trace $\cZ^{fg}{}_f$ 
holds for the same reasons as the vanishing of $\cZ^g{}_f$ did,
namely that $fg$ has vanishing diagonal matrix elements in $\H_f$.
The verification of the traces $ \cZ ^f {}_{fg}$ and $ \cZ ^g {}_{fg}$
will require the construction of the Hilbert space $\H_{fg}$ in the sector 
twisted by $fg$, to be given below. 

\medskip

Finally, it is worth noting that although $fg=gf$ when $f,g$ are viewed
as elements of the point group, they do not commute when they are
viewed as elements of the full orbifold group, since $fgf^{-1} g^{-1}
= (s_L^{-2} , s_R^2)$. For asymmetric orbifold groups, 
we are aware of no arguments based on first 
principles that would guarantee the vanishing of the torus contribution for 
twists $f,g$ that do not commute in the full orbifold group
(but do commute in the point group). Nonetheless, this vanishing
actually does take place in the model considered here. 

\subsubsection{The action of $s_{L,R}$ in the sector twisted by $r_{L,R}$}
\label{sec:shifts}

For the sake of completeness, as well as for later use, we briefly discuss 
the action of $s_L$ on the twisted states $|\pm \>_L$, which enters 
(\ref{gaction}) but is not actually needed for the evaluation of $\cZ^g{}_f$.
In the left-right symmetric conformal field theory of a single scalar field,
twisted by the symmetric reflection operator $r$, the symmetric ground states 
$|\pm \>$ are associated with the fixed  points of $r$. Applying $r$ to the 
circle of radius $R$, there are two solutions  to the 
fixed point equation $r(x) \equiv -x $ (mod $2 \pi R$), 
namely $x=0$ and $x=\pi R$. Under $r$, each fixed point is mapped into itself, 
which is tantamount to $r|+\>\sim |+\>$ and $r|-\> \sim |-\>$, and the state 
$|+\>$ may be associated with the fixed point $0$ while $|-\>$ is associated 
with 
the fixed point $\pi R$. On the other hand, the operator $s$ acts by shifts 
$s(x)=x + \pi R$ and thus  interchanges the two twisted ground states, 
$s|+\> \sim |-\>$ and $s|-\>\sim |+\>$. It is natural and internally consistent  
to induce analogous 
actions on the chiral twisted states, namely $r_L |\pm \>_L \sim |\pm \>_L$ 
and $s_L |\pm \>_L \sim |\mp \>_L$, and $r_R |\pm \>_R \sim |\pm \>_R$ 
and $s_R |\pm \>_R \sim |\mp \>_R$.

\subsubsection{The sector twisted by $fg$}

The only partition trace in the untwisted sector which has not yet been 
determined is  $\cZ^{fg}{}_1(\tau)$. Since the product $fg$ involves reflection
operators on both left and right movers, only states with $n_L=n_R=0$ will 
contribute in the untwisted sector and the partition trace manifestly 
reduces to that of the symmetric operator $r=r_Lr_R$; 
it is readily computed and one finds,
\be
\cZ^{fg}{}_1 =|\zeta[0|{1\over 2}]|^2
\ee
Modular covariance implies
\label{fgpart}
\bea
\cZ ^1 {}_{fg} = 2 |\zeta [\quar |0]|^2
\qquad
\cZ ^{fg} {}_{fg} = 2 |\zeta [\quar |\half]|^2
\eea
The first result yields the Hilbert space of the sector twisted by $fg$ \be
\H_{fg}
=
\oplus
\big\{
x_{-n_1+\half}\cdots x_{-n_k+\half}\tilde x_{-n_1+\half}\cdots
\tilde x_{-n_{\tilde k}+\half}|\pm\>'
\big\}
\ee
Here, $|\pm\>'$ are two  ground states in the sector twisted by
$fg = (r_Ls_L,s_Rr_R)$, both of which are of conformal weight $1/16$.

\medskip

If the presence of the translation operators $s_L$ and $s_R$
were ignored, $fg$ would coincide with the non-chiral reflection $r$,
and $|\pm \>' =|\pm \>$ would just be the ground states in the sector 
twisted by $r$. The problem with this simplified picture is that the vanishing 
of $\cZ ^f {}_{fg}$ does not allow for a {\sl chiral} formulation of the 
states $|\pm \>$ and the operator $fg$  consistent with
the relations $r_L |+\>_L = |+\>_L$ and $r_L |-\>_L = -i |-\>_L$,
used successfully in other sectors, such as (\ref{rLtwisted}), 
(\ref{rL3twisted}), (\ref{rLftwisted}).

\medskip

If the presence of the translation operators $s_L$ and $s_R$ is
carefully taken into account, a consistent chiral formulation of the twisted
states $|\pm \>'$ and the action thereupon by the chiral 
operators $r_{L,R}$ and $s_{L,R}$ does exist and reproduces 
the corresponding partition traces predicted from modular covariance.
The key observation is that the states twisted by $(r_L,r_R)$ and by 
$fg = (r_Ls_L,s_Rr_R)$ are different but isomorphic to one another.
Following the spirit of section \ref{sec:shifts}, the twisted states
may be associated with geometrical fixed points in the symmetric theory.
While the fixed points of $(r_L,r_R)$ on the circle are $0$ and $\pi R$,
those of $fg = (r_Ls_L,s_Rr_R)$ are shifted by $\pi R/2$, i.e. they are 
$+\pi R/2$ and $- \pi R/2$. Denoting associated twisted states by
$|+\>'$ and $|-\>'$ respectively, it is now clear that one should expect to have 
$r |+\>' \sim |-\>'$, $r |-\>' \sim |+\>'$, as well as 
$s |+\>' \sim |-\>'$, $s |-\>' \sim |+\>'$. On the other hand, the states 
$|\pm \>'$ are eigenstates of the operators $rs$.

\medskip

The above geometry-inspired picture may be realized concretely
in terms of chiral states and operators. We start with 
the chiral ground states $|\pm \> _{L,R}$ of the sectors twisted by
$r_L$ and $r_R $ and the action of the chiral reflection operators on these 
states,
\bea
r_L |+\>_L = |+\>_L 
& \qquad &
r_L |-\>_L = -i |-\>_L
\no \\
 r_R |+\>_R = |+\>_R
& \qquad &
r_R |-\>_R = + i |-\>_R
\eea
Using the standard relations $s_L^a r_L s_L^a = r_L$ and 
$s_R^a r_R s_R^a = r_R$ for any $a\in {\bf R}$, one deduces
\bea
\label{rscomps}
(r_Ls_L) |+\>_L' = |+\>_L' 
& \quad &
(r_Ls_L) |-\>_L' = -i |-\>_L'
\hskip .7in
|\pm\>_L' \equiv s_L ^{-\half} |\pm \>_L
\no \\
(s_R r_R) |+\>_R' = |+\>_R'
& \quad &
(s_R r_R) |-\>_R' = + i |-\>_R'
\hskip .67in
|\pm \>_R' \equiv  s_R ^{+\half} |\pm \>_R
\eea
Thus, the action of the operator $r_Ls_L$ on the twisted states $|\pm\>_L'$
is isomorphic to the action of $r_L$ on the twisted states $|\pm \>_L$.
The action of the chiral operators $r_{L,R}$ on $|\pm \>_{L,R}'$, however, 
is now calculable from the above definitions, from $s_L^a r_L s_L^a = r_L$,
and from the fact that $s_{L,R} |\pm\>_{L,R}' \sim |\mp\>_{L,R}'$, and we find,
\bea
r_L |\pm \>_L ' \sim |\mp\>_L'
\qquad \qquad
r_R |\pm \>_R' \sim |\mp\>_R'
\eea
This relation readily guarantees that $f$ and $g$ have vanishing 
diagonal matrix elements in $\H_{fg}$, so that $\cZ ^f {}_{fg} = \cZ^g 
{}_{fg}=0$.

\medskip

The final relation to be implemented is the partition trace $ \cZ ^{fg} {}_{fg}
= 2 |\zeta [\quar|\half]|^2$, which effectively requires that $fg$ be the 
identity
on both states $|\pm \>'$. In view of $fg=(r_Ls_L,s_Rr_R)$ and (\ref{rscomps}),
this provides with a unique correspondence between the chiral
and non-chiral twisted states and we have
\bea
|+ \>' = |+\>_L' \otimes |+\>_R'
\qquad \qquad
|- \>' = |-\>_L' \otimes |-\>_R'
\eea
With this construction, all the partition traces of (\ref{fgpart}) and 
(\ref{fgvanish}) are indeed reproduced in a chiral fashion.
The partition function for the orbifold generated by $f$ and $g$ is,
in dimension $d\in 4{\bf N}$ is therefore given by
\be
Z
={1\over 4}\bigg\{(|\zeta[0|0]|^2+|\zeta[\half|0]|^2)^d+3|\zeta[0|\half]|^{2d}
+
3\cdot 2^d|\zeta[{1\over 4}|0]|^{2d}
+
3\cdot 2^d|\zeta[{1\over 4}|{1\over 2}]|^{2d}\bigg\}
\ee
Notice that this partition function involves the same blocks as
that of the orbifold by $f$ alone, but the relative proportions
of the three non-vanishing modular orbits is different.

\subsection{Including worldsheet fermions}

It is straightforward to include worldsheet fermions 
$\psi ^\mu _\pm$ in models twisted by shifts $s$ and reflections $r$. 
Given a canonical homology basis of 1-cycles $A,B$,
the worldsheet fermions are defined with a spin structure 
$\delta= (\delta '|\delta '')$.
The models of greatest interest are those with worldsheet supersymmetry 
\cite{dgh}. Invariance of the matter supercurrent 
$S_m = -1/2 \psi ^\mu _+ \p_z x^\mu $
thus forces $\psi ^\mu _\pm$ to undergo the same transformation
as the bosonic field $\p_z x$. Shifts do not act on $\psi ^\mu _\pm$.
Reflections may be parametrized by a half-characteristic 
$\ep = (\ep'|\ep'')$, with $ \ep', \ep'' =0,\half$,
such that $\psi ^\mu _\pm$ is double-valued around the 
cycle $D_\ep = \ep ' A + \ep '' B$. The combined boundary
conditions due to the spin structure $\delta$ and the 
reflection $\ep$ are,
\bea
\psi _\pm ^\mu (z+1) & = & - (-)^{2 \delta ' + 2 \ep'} \psi _\pm ^\mu (z)
\\ \no
\psi _\pm ^\mu (z+\tau) & = & - (-)^{2 \delta '' + 2 \ep''} \psi _\pm ^\mu (z)
\eea
The chiral partition function is given by
\bea
\F_\psi [\delta;\ep] = { \tet [\delta + \ep] (0,\tau) \over \eta (\tau)}
\eea
When $\delta + \ep$ equals $(\12 | \12)$, the chiral
partition function vanishes, $\F [\delta;\ep]=0$.

\medskip

Notice that while the bosonic models twisted by $r$ and by $s$ are 
identical, once fermions are included, they will differ, since the 
fermions are twisted by $r$ but not by $s$.

\section{Asymmetric Orbifolds via chiral splitting }
\setcounter{equation}{0}

The operator and Hilbert space methods used earlier in 
this paper to construct the  
partition traces and full partition functions for asymmetric orbifolds
on a torus worldsheet do not easily generalize to higher loop order.
The construction of symmetric orbifolds is, of course, well-understood,
both in the functional integral and the operator formulations 
(see, for example \cite{dfms,g88,dms,DVV,dvvv}).
The construction of the chiral blocks has also been extensively
investigated using operator methods, the chiral OPE, 
and the implications of modular invariance \cite{DVV,dvvv}.
Furthermore, in recent work \cite{adp}, the full $\bZ_2$-twisted chiral blocks 
were calculated in the presence of non-trivial supermoduli
and simple expressions in terms of $\tet$-functions were derived.
It is the construction of the full partition functions for {\sl asymmetric 
orbifolds} that remains much less well-understood.  

\medskip

By chiral splitting, it is
clear that the correct blocks of an asymmetric orbifold
theory should be defined as the products of the blocks 
of the left and right chiral halves. To be concrete, the asymmetric
orbifold group $G$ is viewed as a subgroup of the 
product $G_L \times G_R$ of left and right groups $G_L$ and $G_R$,
so that elements of $G$ may be labeled as pairs $(g_L;g_R)$,
with $g_L \in G_L$ and $g_R \in G_R$. The starting point of 
the proposed construction is two symmetric orbifolds, one with 
group $G_L$, the other with group $G_R$. 
For simplicity, the discussion will be carried out here for the torus,
with boundary conditions around a single $A$ and 
a single $B$ cycle; the generalization to higher genus
are analogous.

\medskip

By chiral splitting, the blocks are given by 
$|\F ^{g_L}{}_{h_L} (p_L;\tau)|^2$ and  
$|\F ^{g_R}{}_{h_R}(p_R;\tau)|^2$ respectively, 
i.e. the left and right chiral blocks are complex conjugates of one another.
Here, $p_L,p_R$ stand for any extra labels of the chiral blocks, 
such as, for example, dependence on the internal loop momentum.
A block $\cZ ^g {}_h$ of the asymmetric orbifold
with $h=(h_L;h_R)$ and $g=(g_L;g_R)$ is then 
defined as the product of the corresponding chiral blocks;
as illustrated by the diagram below,
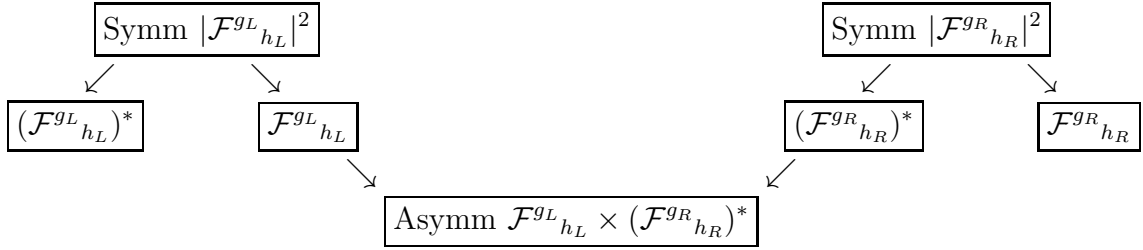
\bea
\fbox{$\displaystyle  {\rm Symm \ } |\F ^{g_L}{}_{h_L}|^2$}
\hskip .2in 
& &
\hskip .2in 
\fbox{$\displaystyle  {\rm Symm \ } |\F ^{g_R}{}_{h_R}|^2$}
\no \\ 
\swarrow \hskip .7in \searrow 
\hskip .4in 
&& 
\hskip .4in \swarrow \hskip .7in \searrow
\no \\
\fbox{$\displaystyle (\F ^{g_L}{}_{h_L})^*$}
\hskip .6in
\fbox{$\displaystyle   \F ^{g_L}{}_{h_L}$}
& \hskip .6in &
\fbox{$\displaystyle   (\F ^{g_R}{}_{h_R})^*$}
\hskip .6in
\fbox{$\displaystyle    \F ^{g_R}{}_{h_R}$}
\no \\ 
 \searrow \hskip -.1in && \hskip -.1in \swarrow 
\no \\
& \fbox{$\displaystyle  {\rm Asymm \ }  \F ^{g_L}{}_{h_L} \times 
(\F ^{g_R}{}_{h_R})^*$}&
\no 
\eea
and may be expressed by the following formula,
\bea
\cZ ^{(g_L;g_R)}{}_{(h_L;h_R)} (p_L,p_R; \tau, \bar \tau)
=
 \F ^{g_L}{}_{h_L} (p_L;\tau) \overline{ \F ^{g_R}{}_{h_R} (p_R;\tau) }
\eea
The key difficulty with asymmetric orbifolds 
resides in the precise pairing between the left and right chiral blocks,
i.e. in the superposition of the blocks $\cZ ^{(g_L;g_R)}{}_{(h_L;h_R)}$.

\medskip

Following \cite{adp}, the superposition of the blocks 
may be expressed in terms of 
a {\sl pairing matrix on left and right chiral blocks}, which is
denoted $K$. The full partition function is then given by\footnote{
It is useful to note that, when considering the compactification
and orbifolding of several dimensions $d>1$, the group
elements $g_L,h_L,g_R,h_R$ represent the action of the orbifold
group on all dimensions and the chiral blocks $\F$ represent the
chiral blocks for $d$ dimensions. When the $d$ dimensions
are orthogonal, as has been the case in this paper, the 
blocks $\F$ themselves are the product of $d$ one-dimensional
blocks and the action of a group element $g$ accordingly
decomposes into $d$ group elements each acting on 
a one-dimensional block.}
\bea
\label{apart}
Z_G (\tau, \bar \tau) 
= 
{1 \over |G|} \sum _{g,h \in G} \ \sum _{p_L,p_R}
\! K(p_L,g_L,h_L;p_R,g_R,h_R) \
\F ^{h_L} {}_{g_L} (p_L;\tau) \left ( \F ^{h_R}{}_{g_R} (p_R;\tau) \right )^*
\eea
Here the summations are over $g=(g_L;g_R)$ and $h=(h_L;h_R)$.
Clearly, the matrix $K$ cannot depend upon moduli $\tau, \bar \tau$
as this would violate chiral splitting. 
Many models admit the same symmetrized theories,
and hence the same chiral blocks. It is then
the pairing matrix $K$ which differentiates between
different models, and the key issue is its determination.

\medskip

Modular invariance places strong constraints on $K$, which
may be derived as follows. The modular transformations of 
the blocks of the symmetric theory are as in (\ref{modularcovariance}).
Using chiral splitting, it follows that the transformations of the 
chiral blocks themselves are known up to phases $\varphi$
and mixing coefficients $\cS$, which are independent of moduli,
\bea
\F ^g {}_h (p_L;\tau+1)  
& = & 
e^{i \varphi (g,h)} \, \F ^{gh^{-1} }{}_h (p_L;\tau) 
\no \\
\F ^g {}_h (p_L;-1/\tau) 
& = & 
\sum _{p_L'} \cS (p_L,p_L')  \F ^{h^{-1} }{}_g (p_L';\tau) 
\eea
A sufficient condition for modular invariance of the full partition
function is the requirement that $K$ satisfy the following relations,
\bea
\label{modK}
K \left (p_L , g_Lh_L^{-1}, h_L;p_R, g_Rh_R^{-1}, h_R \right ) 
& = &
e^{i \varphi (g_L,h_L)-i \varphi (g_R,h_R)} \ 
K \left (p_L,g_L,h_L;p_R,g_R,h_R \right )
\\
K \left ( p_L;h_L^{-1}, g_L;p_R,h_R^{-1}, g_R \right ) 
& = &
\sum _{p_L',p_R'} \cS (p_L,p_L') \cS (p_R,p_R')^*  
K \left ( p_L,g_L,h_L;p_R,g_R,h_R \right )
\no \eea
These conditions are also necessary when all the chiral block-functions
$\F ^g {}_h$ are linearly independent. If the block-functions exhibit 
non-trivial linear dependences, the matrix $K$ can be reduced
to pair only the linearly independent bock-functions; modular 
invariance may then be expressed on this reduced $K$ matrix
just as in (\ref{modK}).

\medskip

Modular invariance cannot, in general, determine $K$ completely,
even when all the block-functions are linearly independent.
The action of modular transformations 
$\tau \to (a \tau + b)  (c \tau +d)^{-1}$ (with $a,b,c,d \in \bZ$ and $ad-bc=1$)
induces the following transformations on pairs of group elements 
\bea
(g_L,h_L)  \to  (g_L^a h_L^b , g_L ^c h_L^d)
\hskip 1in
(g_R,h_R)  \to  (g_R^a h_R^b , g_R ^c h_R^d)
\eea
Generally, this action will decompose into a several disjoint modular orbits.
Clearly, the modular transformation laws of $K$ in (\ref{modK}) cannot 
serve to relate $K$ on different orbits; this information 
must come from physical input, derived from the  
Hilbert space formulation and the proper actions of the 
chiral group elements $g_L,h_L$. 

\medskip

Since  $K$ is independent of moduli, its value may be determined 
by taking various degeneration limits of moduli space in which 
the corresponding chiral blocks  do not all vanish.
For example, in the case of genus two,  the separating degeneration
will produce two tori with prescribed twist sectors.
On each of these tori, the matrix $K$ is known.
Therefore, the genus two matrix $K$ will be known
for the subset of twists for which the genus one 
limits of the chiral blocks are non-vanishing
(and linearly independent).
We expect that the matching of the Hilbert space
construction with the expression of (\ref{apart}) 
produces a unique pairing matrix $K$ to higher genus.

\bigskip\bigskip

\noindent
{\Large \bf Acknowledgements}

\medskip

We are happy to acknowledge helpful conversations with 
Constantin Bachas, Michael Dine, Michael Gutperle, and Per Kraus. 
We are especially grateful to Shamit Kachru and Eva Silverstein
for helpful discussions and correspondence 
on asymmetric orbifolds.

\begin{appendix}
\setcounter{equation}{00}
\section{Orbifolding by symmetric orbifold groups}
\setcounter{equation}{0}

In this appendix, some concrete orbifold partition functions
for symmetric orbifolds are constructed. The general prescription 
in terms of conjugacy classes is reviewed. Finally, simple examples
of symmetric orbifolds based on shifts and reflections are worked out.

\subsection{The functional integral and conjugacy class  prescription}

Consider a scalar field $x$ with classical action $S[x]$
on a torus worldsheet with modulus $\tau$
subject to the following twisted boundary conditions ${\cal B}$ on the 
$A$- and $B$-cycles, corresponding to $z\to z+1$ and $z \to z+\tau$ 
respectively,
\be
{\cal B} \left \{ \matrix{x(z+1)=h\,x(z) \cr 
x(z+\tau)=g\,x(z) \cr} \right .
\ee
The functional integral with boundary conditions ${\cal B}$
is denoted by  $\cZ ^g {}_h(\tau)$ and given by
\bea
\cZ ^g {}_h (\tau) \equiv \int _{\cal B} Dx e^{-S[x]}
\eea
It is well-known that when $gh \not= hg$, the above boundary conditions
have no solutions, and such sectors do not contribute to the 
functional integral. Since the boundary conditions have also the 
following two equivalent representations,
\bea
{\cal B} 
\ \Longleftrightarrow \
\left \{ \matrix{x(z+1)=h\,x(z) \cr 
x(z+\tau +1 )=gh\,x(z) \cr} \right .
\ \Longleftrightarrow \
\left \{ \matrix{x(z+\tau)=g\ x(z) \cr 
x(z-1 )=h^{-1}\,x(z) \cr} \right .
\eea
the following modular identities immediately result from the 
functional integral representation,
\bea
\cZ ^g {}_h (\tau) = \cZ ^{gh} {}_h (\tau+1) = \cZ ^{h^{-1}} {}_g (-1/\tau)
\eea
which, in turn, are equivalent to the modular transformation laws
of (\ref{modularcovariance}). 

\medskip

The general prescription for the one-loop orbifold partition function $Z_G$
defined by the group $G$ is given in terms of a summation 
over all possible twisted boundary conditions $g,h \in G$. The following 
two formulas are equivalent, 
\bea
  \label{orbdef}
  Z_G={1\over|G|}\sum_{g,h\in G\atop hg=gh} \cZ^g {}_h
  = \sum_i{1\over|N_i|}\sum_{h\in N_i} \cZ^{C_i}{}_h
\eea
The second expression above is in terms of a summation over 
all conjugacy classes of $G$, which are indexed by $i$ and 
any representative is denoted by $C_i$ in the above formula. 
Also, $N_i$ is the stabilizer of $C_i$, i.e. the subgroup of all $h\in G$
such that $g$ commutes with every element of $C_i$. 
The equivalence of both expressions is readily established
by using the fact that $  \cZ^g {}_h=\cZ^{ugu^{-1}} {} _{uhu^{-1}}$
for all $g,h,u\in G$, and by breaking up the sum over all $g\in G$
into a sum over $g'$ and $u$ with $g=ugu^{-1}$.

\subsection{The Poisson resummation formula}

The reformulation of momentum and winding mode summations
is carried out using the Poisson resummation formula.
Let $A$ be an invertible $n \times n$ symmetric matrix
and $B,C$ two $n$-column vectors,
\bea
\label{Poisson}
&&
\sum _{m^1, \cdots, m^n \in \bZ}
\exp \left \{
- \pi A_{ij} (m^i + C^i) (m^j + C^j) + 2 \pi i B_i (m^i +C^i) \right \}
 \\ && \qquad
= 
{1 \over (\det A)^\half}
\sum _{m_1, \cdots, m_n \in \bZ}
\exp \left \{
- \pi (A^{-1})^{ij} (m_i - B_i) (m_j - B_j) + 2 \pi i C^i (m_i  - B_i) \right \}
\no
\eea
The derivation of this formula is standard \cite{dp88}.

\subsection{Symmetric Abelian orbifolds defined by $\bZ_N$ shifts}
 
Consider the symmetric orbifold generated by $\bZ _N$ shifts,
$s x = x+ 2\pi\,R/N$.
In this subsection, the theory $S_R^1/s$ is worked out using the 
conjugacy class prescription and it is verified to coincides with the 
theory $S_{R/N}^1$, as should be expected on geometric grounds.
The prescription of the preceding subsection for this case gives,
\be
Z_{S_R^1/s}={1\over N}
\sum_{g,h=1}^ { s^{N-1}} \cZ^g{}_h
\ee
By definition, $\cZ^1{}_1=Z_{S_R^1}$ is the untwisted partition function. 
 
\medskip

General shifted boundary conditions may be specified by the 
characteristics $\delta= (\delta',\delta'')$ where $\delta',\delta ''=
0, 1/N, 2/N, \cdots (N-1)/N $ and the 
following correspondence with the group elements, 
$h = s^{N \delta'}$ and $g=s^{N \delta ''}$,  
 \bea
x(\sigma^1+1,\sigma^2) &=&
x(\sigma^1,\sigma^2) + 2\pi R ~\delta'  \qquad ({\rm mod} \ 2\pi R)
\no \\
x(\sigma^1,\sigma^2+1) &=&
x(\sigma^1,\sigma^2) + 2\pi R ~\delta''
\qquad  ({\rm mod} \ 2\pi R)
\eea
The associated instanton solutions and action are given by
\bea
x_{m_1,m_2}^{\delta} (\sigma)
& = &
2\pi R \sigma^1 (m_1+\delta ')
+
2\pi R  \sigma^2 (m_2+\delta '')
\no \\
S[x_{m_1,m_2}^{\delta}]
& = &
{\pi R^2\over 2 \lbar ^2\tau_2} \bigg |\tau (m_1+\delta') - (m_2+\delta'')
\bigg |^2
\eea
The field $y(\sigma) \equiv x(\sigma) - x_{m_1,m_2}^{\delta}(\sigma)$ is now a 
doubly periodic scalar function. Its functional integral produces the well-known 
factor produces a factor $\Det ' \Delta = \tau _2 |\eta (\tau)|^4$  from the 
non-zero modes and a factor of $(8 \pi ^2 )^{-\half} 2\pi R/ \lbar= R /\s2 
\lbar$ 
from the zero mode of $y$. Assembling all contributions, one finds, (still with 
$h = s^{N \delta'}$ and $g=s^{N \delta ''}$),
\be
\label{cZ1}
\cZ^g {}_h (\tau)
=
{R\over \sqrt{2 \tau_2} \lbar |\eta (\tau)|^2}
\sum_{m_1,m_2\in {\bf Z}}
\exp \left \{ -
{\pi R^2\over 2 \lbar ^2\tau_2} \bigg |\tau (m_1+\delta') - (m_2+\delta'')
\bigg |^2
\right \}
\ee
Under the modular transformations $\tau \to \tau +1$, 
one has $\delta ' \to \delta '$ and $\delta '' \to \delta '' - \delta'$, i.e.
$h\to h$ and $g \to gh^{-1}$ in accord with (\ref{modularcovariance}).
Under the modular transformations $\tau \to -1/\tau$, 
one has $\delta ' \to \delta ''$ and $\delta '' \to  - \delta'$, i.e.
$h\to g$ and $g \to h^{-1}$, also in accord with (\ref{modularcovariance}).

\medskip

The formula (\ref{cZ1}) obtained above has manifest modular
transformation properties but does not exhibit the chiral block
structure of the theory. To make the chiral block structure manifest,
Poisson resummation in $m_2$ is carried out, using (\ref{Poisson}) for $n=1$,
$A= R^2 /2 \lbar ^2 \tau_2$, $B=0$ and $C=\delta '' - \tau_1 (m_1 + \delta')$, 
\be
\label{cZ2}
\cZ^g {}_h (\tau)
=
{1\over |\eta(\tau)|^2}
\sum_{(p_L,p_R) \in \Gamma ^{\delta'} _R}
e^{2\pi i\delta'' (p_L+p_R) {R \over 2 \lbar}} \
q^{{1\over 2}p_L^2} \
\bar q^{{1\over 2}p_R^2}
\ee
Here,  the momenta $(p_L,p_R)$ span the  lattice 
$\Gamma ^{\delta'}_R$, defined by,
\bea
\Gamma _R ^{\delta'} \equiv \left \{ 
(p_L ,p_R) = \left (
{\lbar \over R}m_2-{R\over 2\lbar }(m_1+\delta') ,
{\lbar \over R}m_2+{R\over 2\lbar }(m_1+\delta')  \right ) 
\quad m_1,m_2 \in \bZ \right \}
\eea
The modular transformation properties of (\ref{cZ2}) of course follow
from those of (\ref{cZ1}) which were discussed after (\ref{cZ1}). But they
may also be derived directly from (\ref{cZ2}). 
This is manifest for $\tau \to \tau+1$, while for $\tau \to \tilde \tau = 
-1/\tau$, 
it is achieved by applying the Poisson resummation formula (\ref{Poisson}) 
in both $m_1$ and $m_2$ for $n=2$ and 
\bea
A= \left ( \matrix{
{R^2 \over 2 \lbar ^2} \tilde \tau _2 & - i \tilde \tau \cr 
-i \tilde \tau & 2{R^2 \over \lbar ^2} \tilde \tau_2\cr} \right )
\hskip .7in
B =\left ( \matrix{0 \cr \delta '' \cr} \right )
\hskip .4in
C =\left ( \matrix{\delta ' \cr 0 \cr} \right )
\eea

and one recovers results in accord with (\ref{modularcovariance}).

\medskip

The summation over $\delta''$ forces $m_2= (p_L+p_R) {R \over 2 \lbar}$ 
to be an integer multiple of $N$, so that $m_2 = N n_2$ with $n_2 \in \bZ$,
and eliminates the overall factor of ${1\over N}$.
Next, the summation over $\delta'$ is reproduced by replacing
$m_1$ with ${1\over N}n_1$, where $n_1\in {\bf Z}$.
Combining all, one finds,
\be
Z_{S_R^1/s}
=
{1\over |\eta(\tau)|^2}
\sum_{(p_L,p_R)\in \Gamma ^0 _{R/N}}
q^{{1\over 2}p_L^2}
\bar q^{{1\over 2}p_R^2}
\ee
This expression clearly coincides with $Z_{S^1{R/N}}$.
The above formula can be understood in the Hamiltonian picture as a trace 
over a Hilbert space $\H _{s^a}$ twisted by $s^a$ with the insertion of
the $b$-th power of the translation operator $s$ of shifts by $2\pi R/N$,
\bea
  \label{eq:zTrace}
  \cZ^g {}_h={\rm Tr}_{s^a}\left( s^b ~ q^{p_L^2/2}\bar
q^{p_R^2/2}\right)
\qquad\quad
s \equiv \exp\left ( 2\pi i(p_L+p_R){R\over 2 \lbar N}\right )
\eea

\section{The point group versus by the full orbifold group}
\setcounter{equation}{0}

In the construction of an orbifold theory of flat space $\bR^n$, 
the orbifold group $G$ (or more generally the space group of $G$)
is a discrete subgroup of the Euclidean group of $\bR^n$,
consisting of elements $g=(R_g,v_g)$, where $R_g \in O(n)$
is a rotation and $v_g \in \bR^n$ a translation. The maximal subgroup
of pure translations is denoted $\Lambda _G$ while the subgroup
of all elements $R_g$ is the point group $P_G$. The normal 
subgroup $\bar P_G \equiv G/\Lambda _G$ is isomorphic to
$P_G$, while the coset $\bR^n/\Lambda _G={\bf T_G}^n$ is 
a torus. Symmetric orbifolds may be constructed either 
coseting $\bR^n$ by the full $G$ or coseting ${\bf T}^n $
by $\bar P_G$,
\bea
\bR^n /G = {\bf T_G}^n/\bar P_G
\eea 
In this appendix, the simplest non-trivial such case 
when $n=1$ and $G= \bZ_2 \times \bZ$ will be 
shown to yield the same partition function when treated
either way.

\subsection{The functional integral from $S^1/{\bf Z_2}$}

Here, the field $x$ takes values in the circle $S^1_R$.
The point group  $\bar P={\bf Z_2} = \{ (1,0), (-1,0)\}$ is
Abelian, and hence each element forms a conjugacy class by itself.
The centralizer of each element is always the full $\bar P$, so
that the cardinality of the centralizer is always 2, producing 
a factor $\half$.
In the twisted sectors, consider all solutions to the twisted
boundary conditions. For example, when the twist is placed on the
$B$-cycle,
\bea
x(\sigma ^1 +1, \sigma ^2) & = &
+ x (\sigma ^1, \sigma ^2) \quad mod (2 \pi R)
\no \\
x(\sigma ^1 , \sigma ^2+1) & = &
- x (\sigma ^1, \sigma ^2) \quad mod(2 \pi R)
\eea
the general solution is a combination of oscillating solutions,
plus a constant,
\bea
x_{m_1, m_2} (\sigma ^1, \sigma ^2) = x_0 +
\exp \bigg \{ 2 \pi i \bigg [ m_1 \sigma ^1 + (m_2 +\half) \sigma
^2 \bigg ] \bigg \}
\eea
The constant may be interpreted as the center of mass of the string, as
usual. It must satisfy $x_0 = - x_0 \quad mod(2 \pi R)$, which produces
two solutions or {\sl fixed points} $x_0 =0$ and $ x_0 = \pi R$.
Both fixed points produce equal contributions to the functional
integral, whence a factor of 2. Putting all together, one finds
\bea
\int Dx \ e^{-S[x]} = \half Z_{S^1_R} + \sum _{i=2,3,4} \left | {\eta
(\tau ) \over \tet _i (0,\tau) } \right |
\eea
This formula agrees with \cite{g88}.

\subsection{The functional integral from ${\bf R}/({\bf Z_2} \times {\bf Z})$}

Quotienting by the full orbifold group, the conjugacy classes are

\bea
C_1  = \{ (1,0) \}             &\qquad & N_1 = G \no \\
C_2  = \{ (1,m), \ m >0 \}& \qquad  & N_2 = \{ (1,n), \ n \in \bZ \} \no \\
C_3  = \{ (-1,0) \}            &\qquad  & N_3 = \{ (1,0), (-1,0) \} \no \\
C_4  = \{ (-1,0) \}            & \qquad & N_4 = \{ (1,0), (-1,1) \} 
\eea
For the classes $C_3$ and $C_4$, the centralizers have cardinality
2, resulting in an overall factor of 1/2 in the partition
function. On the other hand, the contributions from $C_3$ and
$C_4$ are identical, therefore cancelling the factor of 1/2. Their
contributions result in the twisted functional integrals with
$\tet _i$ for $i=3,4$. For the classes $S_1$ and $S_2$, the cardinality of the
centralizers are
\bea
\# (N_1) = 2 \#( \Lambda_R )
\qquad \quad
\# (N_2) = \#( \Lambda _R )
\eea
Recall that $\Lambda _R =\{ 2 \pi n R , \ n\in {\bf Z} \}$. The factor of 2
arises because $N_1 = {\bf Z}_2 \times \Lambda _R$.

\medskip

The elements in $N_1$ of the type $(-1,n)$ produce $\#(\Lambda _R )$
identical copies of the twisting $(-1,0)$, which yields the
contribution with $\tet _2$. To derive the precise weight of this
contribution is a little tricky. For a given element $(-1,n)$,
the boundary conditions are
\bea
x(\sigma ^1 + 1, \sigma ^2) & = & +x(\sigma ^1, \sigma ^2)
\no \\
x(\sigma ^1 , \sigma ^2+1) & = & - x(\sigma ^1, \sigma ^2) + 2 \pi n R
\eea
The general solution of these equations is
\bea
x_{m_1,m_2} (\sigma ^1 , \sigma ^2) = \pi n R + e^{2 \pi i ((m_1+\half)
\sigma ^1 + m_2 \sigma ^2)}
\eea
Here, $n \in \bZ$ specifies the center of mass of the string at $\pi nR$.
These positions may be viewed as translates of one another by the lattice 
$\Lambda _{\half R}$. The original shifts in the boundary conditions were
$2\pi nR$, which may be viewed as translates in the lattice $\Lambda _R$.
Although both quantities are infinite, their ratio is well-defined, and given by
$\# (\Lambda _{\half R} ) = 2 \# (\Lambda _R)$.
As a result, the factor of $\# (\Lambda _{\half R} )$ cancels the factor
of $1 / \# (N_1)$ and the contribution involving $\tet _2$ is recovered
with coefficient 1.
The elements in $N_1$ of the type $(1,n)$ combine with all of
$N_2$ to produce half of the untwisted partition function.

\section{$\tet$-function identities}
\setcounter{equation}{0}

Throughout, the following correspondence of $\tet$-function 
notations is being used, (see \cite{Bat}),
\bea
\tet _1 (z,\tau) = \tet [\12|\12](z,\tau)
& \qquad &
\tet _3 (z,\tau) = \tet [0|0](z,\tau)
\no \\
\tet _2 (z,\tau) = \tet [\12|0](z,\tau)
& \qquad &
\tet _4 (z,\tau) = \tet [0|\12](z,\tau)
\eea
where the $\tet$-functions with half-characteristics are defined by,
\bea
\tet [\alpha |\beta ](z,\tau) \equiv
\sum _{n \in \bZ} e^{i \pi \tau (n + \alpha)^2 + 2 \pi i (n + \alpha) (z + 
\beta)}
\eea
The $\tet$-constants are defined by setting $z=0$. The Dedekind $\eta$-function
is defined by
\bea
\eta (\tau) \equiv e^{i \pi \tau /12} \prod _{n=1} ^\infty (1 - e^{2 \pi i n 
\tau})
\eea
and satisfies $2 \eta (\tau)^3 = \tet_2 (0,\tau) \tet_3 (0,\tau) \tet_4 
(0,\tau)$.
The modular transformations of the $\tet$-constants  and $\eta$-function 
are given by
\bea
\label{theta1}
\tet _2 (0,\tau+1)= e^{i\pi/4} \tet _2 (0,\tau)
& \qquad &
\tet_2 (0,-1/\tau) = \sqrt{-i \tau} ~\tet _4 (0,\tau)
\no \\
\tet _3 (0,\tau+1)=  \tet _4 (0,\tau)
\hskip .32in
& \qquad &
\tet_3 (0,-1/\tau) = \sqrt{-i \tau} ~\tet _3 (0,\tau)
\no \\
\tet _4 (0,\tau+1)=  \tet _3 (0,\tau)
\hskip .32in
& \qquad &
\tet_4 (0,-1/\tau) = \sqrt{-i \tau} ~\tet _2 (0,\tau)
\no \\
\eta (\tau+1) = e^{i \pi /12} \eta (\tau)
\hskip .32in
& \qquad &
\eta (-1/\tau) = \sqrt{-i\tau} ~\eta (\tau)
\eea
The Jacobi $\tet$-identity is 
$\tet_2(0,\tau)^4 + \tet _4(0,\tau)^4 = \tet _3 (0,\tau)^4$.
As a result, the following combinations of $\tet^4$ may be expressed 
in terms of 8-th powers of $\tet$,
\bea
\label{theta2}
2 \tet _2^4 \tet _3^4 & = & \tet _2 ^8 + \tet _3 ^8 - \tet _4^8
\no \\
2 \tet _3^4 \tet _4^4 & = & \tet _3 ^8 + \tet _4 ^8 - \tet _2^8
\no \\
-2 \tet _4^4 \tet _2^4 & = & \tet _2 ^8 + \tet _4 ^8 - \tet _3^8
\eea
The following {\sl doubling identities}, which relate 
$\tet$-constants with modulus $2\tau$ to those of modulus $\tau$,
will be needed. For general characteristics,
\bea
\label{doubling1}
\tet [\alpha |\beta] (0,2 \tau) ^2
=
\half \biggl ( \tet [0|0] ~ \tet [2 \alpha |\beta]
+  e^{-2\pi i \alpha} \tet [0|\12 ] ~ \tet [2 \alpha |\beta + \12]
\biggr ) (0,\tau)
\eea
For some of the characteristics needed here, for example,
\bea
\label{quarttheta}
\tet [{1 \over 4}|0](0,2\tau)^2 = \half \tet _2 \tet _3 (0,\tau)
& \qquad &
\tet [{3 \over 4}|0](0,2\tau)^2 = \half \tet _2 \tet _3 (0,\tau)
\no \\
\tet [{1 \over 4}|\half ](0,2\tau)^2 = + {i \over 2} \tet _2 \tet _4 (0,\tau)
& \qquad &
\tet [{3 \over 4}|\half ](0,2\tau)^2 = - {i \over 2} \tet _2 \tet _4 (0,\tau)
\eea

\subsection{The function $\zeta[\alpha|\beta](\tau)$}

The definition of the function $\zeta [\alpha |\beta](\tau)$ 
may be given either in terms of $\tet$-functions for 
modulus $2 \tau$, or directly in terms of momentum summations,
(with $q\equiv e^{2 \pi i \tau}$),
\bea
\zeta [\alpha | \beta ] (\tau)
\equiv
e^{ - 2 \pi i \alpha \beta } {\tet [\alpha | \beta ] (0,2 \tau) \over \eta 
(\tau)} 
=
 {1 \over \eta (\tau)} \sum _{n\in \bZ} q^{(n+\alpha)^2}
e^{ 2 \pi i n \beta  }
\eea
The following periodicity properties are readily derived,
\bea
\label{zeta0}
\zeta [-\alpha |-\beta ] (\tau) & = & \zeta [\alpha |\beta ](\tau)
\no \\
\zeta [\alpha +1 | \beta ] (\tau) & = &
e^{-2\pi i \beta} \zeta [\alpha |\beta ] (\tau)
\no \\
\zeta [\alpha  | \beta +1 ] (\tau) & = &
 \zeta [\alpha | \beta ] (\tau)
\eea
Modular transformations act as follows
\bea
\zeta [\alpha  | \beta ] (\tau+1) & = &
\omega ~ e^{ 2 \pi i \alpha ^2} \zeta [\alpha | \beta +2 \alpha ] (\tau)
\no \\
\zeta [\alpha  | \beta ] (-{1 \over \tau}) & = &
{1 \over \s2} e^{-2\pi i \alpha \beta}
\biggl (\zeta \biggl [-{\beta \over 2} |2 \alpha \biggr ] (\tau) 
+
e^{ 2 \pi i \alpha}
\zeta \biggl [\half -{\beta \over 2} |2 \alpha \biggr  ] (\tau)
\biggr )
\eea
where $\omega = \exp \{ -i \pi /12\}$. Actually, it is useful to
spell out the modular transformations on the characteristics 
that are needed in the orbifold constructions in this paper,
\bea
\label{zeta1}
\zeta [0|0] (\tau +1 ) =
\omega ~ \zeta [0|0] (\tau  )
& \qquad &
\zeta [0|0] (-1/\tau) ={1 \over \s2} \biggl (\zeta [0|0] + \zeta [\half |0] 
\biggr ) (\tau)
\no \\
\zeta [0|\half] (\tau +1 ) =
\omega ~ \zeta [0|\half ] (\tau  )
& \qquad &
\zeta [0|\half ] (-1/\tau) =
\s2 \zeta [\quar |0] (\tau)
\no \\
\zeta [\half |0] (\tau +1 ) =
\omega ~e^{i \pi /2} \zeta [\half |0] (\tau  )
&\qquad &
\zeta [\half |0] (-1/\tau) =
{1 \over \s2} \biggl (\zeta [0|0] - \zeta [\half |0] \biggr ) (\tau)
\no \\
\zeta [\quar|0] (\tau +1 ) =
\omega ~e^{i\pi /8} \zeta [\quar|\half] (\tau  )
& \qquad &
\zeta [\quar |0] (-1/\tau) =
{1 \over \s2} \zeta [0|\half]  (\tau)
\no \\
\zeta [\quar|\half ] (\tau +1 ) =
\omega ~e^{i\pi /8} \zeta [\quar|0] (\tau  )
& \qquad &
\zeta [\quar |\half ] (-1/\tau) =
\zeta [\quar |\half ]  (\tau)
\eea
Another useful fact is that for $a\in 2\bZ +1$, 
\bea
\label{zeta2}
\zeta [{a \over 4}+\half |{b \over 2}] = 
e^{i\pi (a+1)b/2} \zeta [{a \over 4} |{b \over 2}]
\eea
both of which are proportional (with a $\pm$ factor) to $\zeta
[\quar|{b \over 2}]$. Other useful identities are,
\bea
{q^{-1/24} \over \prod_{n=1}^{\infty}(1+q^n)}
=
\zeta[0|{1\over 2}] (\tau)
\hskip 1in
{q^{1/48}\over\prod_{n=1}^{\infty}(1-(-)^b q^{n-\half})}
= 
 \zeta[{1\over 4}|{b \over 2}](\tau) 
\eea

\end{appendix}

\end{document}